\documentclass{article}

\usepackage{arxiv}

\usepackage[utf8]{inputenc} 
\usepackage[T1]{fontenc}    
\usepackage{hyperref}       
\usepackage{url}            
\usepackage{booktabs}       
\usepackage{amsfonts}       
\usepackage{nicefrac}       
\usepackage{microtype}      
\usepackage{lipsum}		
\usepackage{graphicx}
\usepackage[square,sort&compress,numbers]{natbib}
\usepackage{doi}
\usepackage{braket}
\usepackage{tabularx}
\usepackage{amsmath}
\usepackage{mathtools}
\usepackage{xfp}
\usepackage{siunitx}

\newcommand{\angfreq}[1]{%
  \num[round-mode=places,round-precision=5]{\fpeval{2*pi*(#1)}}%
}

\title{Phase Singularities as a Route to Partial Incoherence in Coupled Oscillators -- From Experiment to Theory}
\usepackage{authblk}

\author[1]{Nicolas Thomé}
\author[1]{Yukiteru Murakami}
\author[1]{Yannick Schöhs}
\author[1\thanks{\tt{Krischer@tum.de}}]{Katharina Krischer}

\affil[1]{School of Natural Sciences, Physics Department, Nonequilibrium Chemical Physics, Technische Universität München, James-Franck-Str. 1, D-85748 Garching, Germany}

\hypersetup{
pdftitle={A template for the arxiv style},
pdfsubject={},
pdfauthor={David S.~Hippocampus, Elias D.~Striatum},
pdfkeywords={First keyword, Second keyword, More},
}

\begin{document}
\maketitle

\begin{abstract}
Phase-only models have contributed significantly to the understanding of synchronization. However, they do not account for dynamical scenarios where amplitude dynamics matter. This study identifies an amplitude-mediated transition from complete frequency coherence to partial incoherence, observed both in an electrochemical silicon-etching experiment and within a population of globally coupled heterogeneous Stuart–-Landau oscillators. Under strong global coupling, the oscillators self-organize into a bimodal-amplitude distribution, from which the transition to partial incoherence is triggered by successive secondary Hopf bifurcations in low-amplitude oscillators. When these modulations cause oscillators to experience momentary vanishing amplitudes, the winding number changes, converting the secondary frequency into a new, oscillator-specific mean frequency. This mechanism results in a partially incoherent state, in which one amplitude group maintains frequency locking while another develops a dispersed frequency branch. These findings demonstrate that phase singularities offer a pathway to incoherence that phase-only models cannot capture.

\end{abstract}


\section{\label{sec:Intro}Introduction}
The spontaneous emergence of coherence in heterogeneous populations of oscillators is a central problem in nonlinear dynamics. Following experimental observations of synchronization in nature~\cite{Buck.1935}, Winfree~\cite{Winfree.1967} introduced one of the first theoretical descriptions of collective synchronization using a phase-reduction approach. Kuramoto~\cite{Kuramoto.1975} subsequently formulated an analytically tractable mean-field model and determined the onset of synchronization in a heterogeneous population of self-sustained oscillators.

Building on these results, the transition from incoherence to coherence has been studied extensively in phase-oscillator networks~\cite{Strogatz.2000}, including the effects of coupling topology and intrinsic-frequency distributions~\cite{Acebron.2005}. The phase-only description, however, is generally obtained under the assumption that coupling and heterogeneity are weak compared to the transverse relaxation toward the limit cycle~\cite{Winfree.1967, Kuramoto.1984, Pikovsky.2010}. In this regime, amplitude deviations remain small and can be eliminated through phase reduction. Outside this regime, amplitude dynamics may influence collective behavior and must be explicitly retained.

A model incorporating both amplitude and phase dynamics is the Stuart--Landau model. Therefore, a heterogeneous population of globally coupled Stuart–-Landau oscillators constitutes a natural model for investigating collective dynamics beyond the phase approximation. In the weak-coupling limit, the first-order phase reduction maps the heterogeneous Stuart–-Landau system onto a Kuramoto–Sakaguchi model~\cite{Kuramoto.1984,Sakaguchi.1986, Pikovsky.2010}. For a unimodal frequency distribution and effectively attractive coupling, this phase-reduced model exhibits a continuous transition to synchronization. In contrast, the analogous heterogeneous Stuart--Landau system with reactive coupling can exhibit a discontinuous transition to synchrony, demonstrating that such a transition can be mediated purely by amplitude dynamics~\cite{Cross.2004,Wang.2016}. 
Extending the phase reduction to the second order can capture some features of the heterogeneous Stuart--Landau system that do not emerge in the first-order approximation. It has been reported that multistability as well as quasiperiodic and chaotic states also occur in the second-order phase reduction~\cite{Leon.2022}. Nevertheless, the effective coupling in this scenario remains in the weak-coupling regime.

When the coupling becomes sufficiently strong, the assumptions underlying standard weak-coupling phase reduction are no longer satisfied~\cite{Nakao.2016}. For example, phase-only models cannot explicitly represent the vanishing oscillation amplitudes considered here. In particular, the amplitude can oscillate and momentarily vanish, a feature known as a phase singularity~\cite{Winfree.2001}. When an experimental setup is considered and spatial diffusion cannot be ignored, a zero in the complex field is also known as an amplitude defect~\cite{Mikhailov.1999}, or, in the case of a nonzero spatial winding number, as a topological turbulence~\cite{Aranson.2002}. Prominent experimental examples underscore the importance of amplitude dynamics in interacting oscillatory systems ~\cite{Varela.2005, Beta.2006, Patzauer.2021} . For systems without spatial extension, it has furthermore been shown that the additional amplitude degree of freedom can enable synchronization by an amplitude adaptation~\cite{Yamaguchi.1981,Yamaguchi.1984}. Previous studies have established that amplitude dynamics, in globally coupled systems, can modify the onset of synchronization and generate amplitude-modulated, frequency-synchronized or more complex collective states~\cite{Cross.2004,Wang.2016,Leon.2022,Matthews.1991,Aronson.1990}. What remains less well understood is how amplitude dynamics, or, more specifically, vanishing amplitudes, can mediate the loss of complete frequency locking in a large, heterogeneous population. Specifically, the dynamical mechanism by which an amplitude modulation is converted into a persistent change of an oscillator's mean frequency remains unexplored.

We employ the anodic electrochemical dissolution of silicon as an experimental model system for investigating the dynamics of coupled oscillators. 
In fluoride-containing electrolytes under anodic polarization, silicon dissolution undergoes spontaneous oscillations in current and oxide-layer properties rather than reaching a stationary state \cite{Turner.1958}. 
These oscillations result from nonlinear feedback between two interfacial reactions that jointly control oxide growth and etching. 
Consequently, the local dissolution dynamics evolve, mediating both electronic and chemical coupling across the electrode surface. For n-type silicon electrodes, the resulting oxide patterns can be systematically manipulated by varying the photoexcitation intensity. Here, we introduce an amplitude-mediated transition from synchrony to partial incoherence that arises in experiments involving an oscillatory, continuous medium with strong global coupling and weak diffusive coupling. The experimental system transitions from a bimodal-amplitude solution with a homogeneous dominant frequency~\cite{Thome.2026} to a solution with one locked frequency and many drifting dominant frequencies. We elucidate the mechanism of the transition using a system of globally coupled heterogeneous Stuart--Landau oscillators. We show that the oscillators in the low-amplitude mode undergo successive Hopf bifurcations, thereby introducing secondary frequencies into the system. These secondary frequencies are then incorporated into the dominant frequency via a phase singularity. 

The manuscript is organized as follows: In section~\ref{sec:Results}, we first present the experimental results and then numerically reproduce them. At the end of the section, we identify a novel transition to partial incoherence, mediated by phase singularities. In section~\ref{sec:Discussion}, we discuss the introduced mechanism by distinguishing it from the well-known phase-only model, comparing it to previous works, and generalizing its applicability. In chapter~\ref{sec:Methods}, we introduce the experimental setup and define all the mathematical concepts used in the results manuscript.

\section{\label{sec:Results} Results}
\paragraph{Eletrochemical Dissolution of Illuminated n-Type Silicon}
The electrochemical system contains two distinct sources of long-range coupling. The external series resistance couples the local dynamics through the total current flowing across the electrode. In addition, illumination controls the generation of valence-band holes, which are required for the oxidation reaction. At sufficiently low illumination, the finite supply of photogenerated holes becomes limiting, causing different regions of the electrode to compete for the available charge carriers. The characteristic lifetime of these minority carriers is several orders of magnitude shorter than the electrochemical oscillation timescale, while their diffusion length is comparable to the lateral dimensions of the electrode~\cite{Schoenleber.2014}. Consequently, redistribution of the charge carriers occurs on a timescale much shorter than the local electrochemical dynamics, motivating an effectively global description of this coupling. 

\begin{figure*}[h!]
\centering
\includegraphics[width=1.0\textwidth]{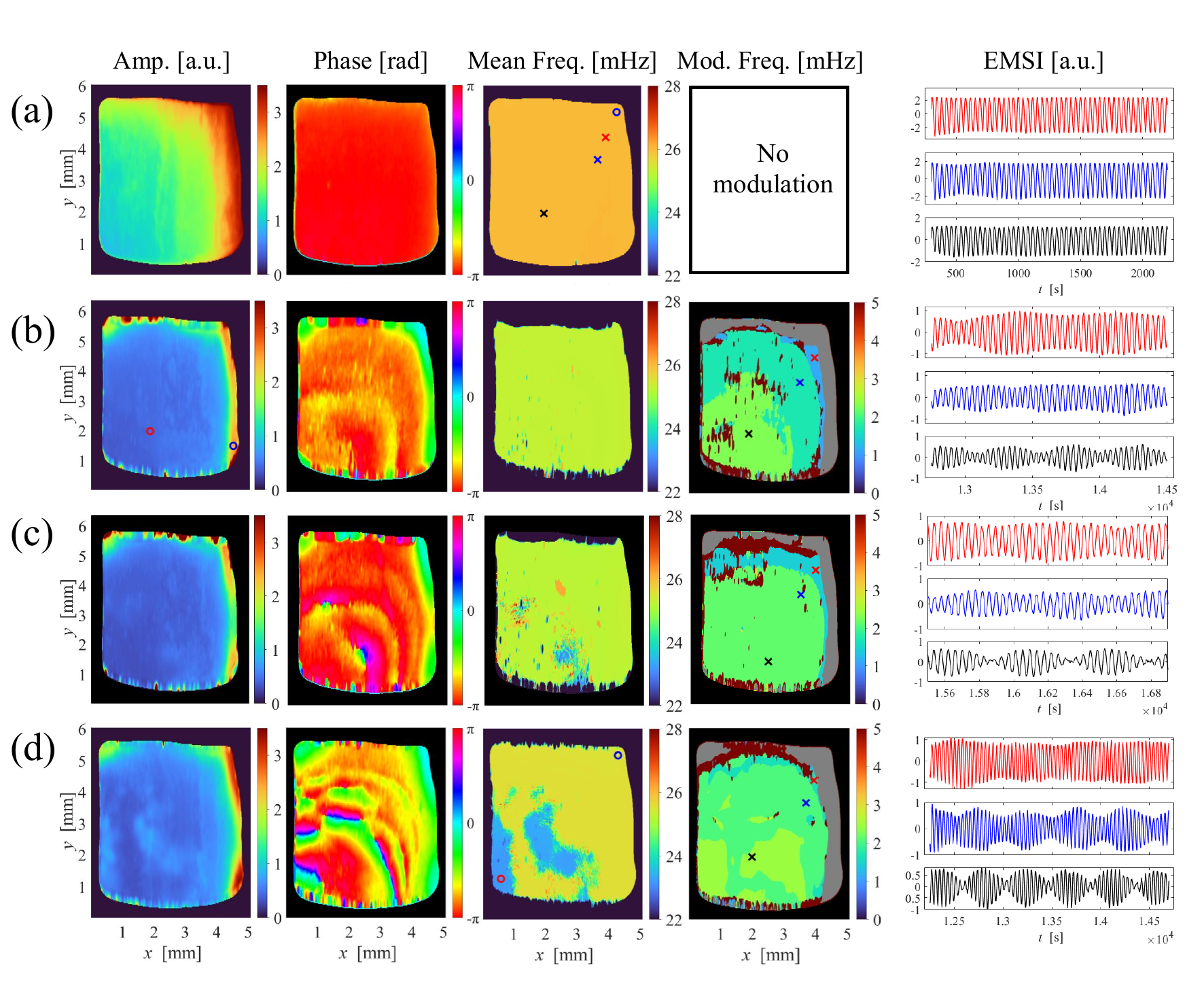}
\caption{\textbf{Experimental Heatmaps:}
Experimentally measured spatial distribution of time-averaged amplitude (first column), representative snapshots of phase distribution (second column), and time-averaged frequency distribution (third column), amplitude-modulation frequency (fourth column) on an electrode surface, and local EMSI signals at three representative points marked with red, blue, and black crosses (fifth column). Circular markers indicate locations we use for FFT analysis in Figure~\ref{fig:Fig_4}. From (a) to (d), the illumination intensity was decreased in steps:
(a) $\mathrm{3.13~mW/cm^2}$,
(b) $\mathrm{0.94~mW/cm^2}$,
(c) $\mathrm{0.89~mW/cm^2}$,
(d) $\mathrm{0.85~mW/cm^2}$.
Gray regions in the fourth column indicate locations where no clear amplitude modulation could be identified.
}
\label{fig:Fig_1}
\end{figure*}

Figure~\ref{fig:Fig_1} shows the evolution of the system states, labeled (a)–(d), observed during the stepwise decrease of the illumination intensity at $3.13$, $0.94$, $0.89$ and $0.85~\mathrm{mW/cm^2}$, while the second global coupling parameter is held constant. 
The first three columns display the results from ellipsomicroscopy for surface imaging (EMSI) measurements across the entire electrode surface. In particular, we can extract phase $\phi(\vec{x},t)$ and amplitude $A(\vec{x},t)$ from the EMSI signal $\xi(\vec{x},t)$ (cf. eq.~\eqref{eq:Hilbert_Tr}) and obtain the time-averaged amplitudes, phase snapshots, and the time-averaged dominant-frequency.

Under the highest illumination intensity, state (a), complete frequency synchronization is observed, and the frequencies are spatially homogeneous.
In contrast, the amplitude and phase distribution are spatially inhomogeneous. The highest amplitudes are located in the upper-right corner and gradually decrease toward the lower-left corner. For the phases, a similar, but much less pronounced gradient is present. The observed spatial inhomogeneity of the amplitudes originates from intrinsic heterogeneities across the silicon electrode surface. 

In state (b), the amplitude variation becomes more pronounced, and distinct high- and low-amplitude regions spontaneously emerge.
The high-amplitude regions are located at the right edge of the electrode and in the upper-left corner, whereas the remainder of the electrode oscillates with a lower amplitude.
While the phase distribution of the low-amplitude region shows slight variations, that of the high-amplitude regions remains homogeneous, and the phase leads that of the low-amplitude region. The frequency is spatially uniform across the entire electrode.

In state (c), the time-averaged amplitude profile remains similar to the previous state. The mean frequencies reveal two distinct spatial regions. Besides the region with a frequency similar to (b), two areas exhibit lower frequencies than in (b). In those spatial regions, the phase snapshots range from $-\pi$ to $\pi$. The rest of the phase profile looks similar to state~(b).

In state (d), the low-frequency regions spread in the bottom-left corner and the central part of the electrode. The time-averaged amplitude distribution remains similar to that of the previous state (c), although the low-amplitude region exhibits stronger amplitude variations. In the phase snapshot, a traveling-wave pattern is observed in the low-amplitude region, whereas the high-amplitude regions at the upper-left corner and the right edge of the electrode remain fully phase-locked.

The fifth column of Fig.~\ref{fig:Fig_1} shows the EMSI signal $\xi(\vec{x},t)$ of three representative points marked by crosses in red, blue, and black.
In state (a), the amplitude of the oscillation is nearly constant at all three points.
On the other hand, in states (b), (c), and (d), an amplitude modulation is observed.
These modulations are associated with the Hopf frequency reported by Patzauer et al.~\cite{Patzauer.2021}.
The modulation at the red points is only weakly visible, and the modulation period is the longest. In the black points, the modulation is most prominent, and the modulation period is the shortest. Specifically, for (c), the amplitude of the oscillation in the EMSI signal vanishes at specific time moments. The amplitude and period of the modulation at the blue points are between those of the black and red points. The local EMSI signals suggest that larger modulation amplitudes tend to occur at locations with higher modulation frequencies.

The fourth column shows the amplitude-modulation frequency distribution, obtained from a Fourier analysis of the local analytic signals.
In state (a), no modulation is observed anywhere on the electrode, as seen in the fifth column.
Also in states (b), (c), and (d), the gray regions correspond to areas where amplitude modulation cannot be clearly identified. In contrast, the remaining parts of the electrode exhibit modulation frequencies, represented by the differently colored regions. In the dark-red regions (frequency larger than $\mathrm{5~mHz}$), the peaks of the amplitude modulation are not well defined in the Fourier spectra, suggesting that the observed variations are likely dominated by noise. In this measurement, the frequency resolution of the Fourier spectra is $\Delta f = 0.57~\mathrm{mHz}$ in (b), $\Delta f = 0.71~\mathrm{mHz}$ in (c), and $\Delta f = 0.41~\mathrm{mHz}$ in (d). The differently colored regions are separated by frequency intervals of $\Delta f$, whereas the actual modulation frequency is expected to vary continuously across the electrode. The lower frequency resolution in~(c) may account for the appearance of only two distinct frequency regions, since continuously varying modulation frequencies can become indistinguishable within individual frequency bins.

Furthermore, comparison of the first and fourth columns reveals that lower amplitudes tend to be associated with higher amplitude-modulation frequencies. In state (d), the shapes of the high modulation frequency regions resemble those of the low-frequency domains in the third column, as well as an increase in the mean amplitude in the first column. This observation indicates a spatial correspondence between modulation frequency and the mean-frequency domains observed in state (d). While this correspondence is initially empirical, the numerical analysis below reveals how an amplitude modulation can mediate a change in mean frequency through a phase singularity, providing a mechanistic explanation for the experimentally observed relationship.

\paragraph{Coupled Amplitude-Oscillator Model}
To investigate the mechanism underlying the experimentally observed separation into distinct dominant frequencies, we employ globally coupled heterogeneous Stuart--Landau oscillators. The Stuart--Landau equation provides a generic amplitude description of oscillations arising from a Hopf bifurcation and therefore constitutes a minimal model that captures the relevant amplitude dynamics of the electrochemical system, as discussed in detail in Ref.~\cite{Thome.2026}. Neglecting spatial diffusion, it corresponds to the spatially uncoupled limit of the heterogeneous complex Ginzburg--Landau model. The experiment contains two physically distinct sources of global coupling: electrical feedback through the external series resistance and carrier-mediated coupling through the limited supply of photogenerated holes. Since these mechanisms act through different physical variables (electric potential and electrochemical potential of the valance band holes) and their microscopic projection onto the oscillatory dynamics is unknown, we phenomenologically represent their combined effect as a linear complex mean-field coupling, following Ref.~\cite{Thome.2026}. Its dissipative and reactive contributions are parameterized by $K$ and $KC_1$, respectively.

\begin{align} \partial_t W_k &= W_k(1+\mathrm{i}\omega_k) -(1+\mathrm{i}C_2)\lvert W_k\rvert^2W_k +K(1+\mathrm{i}C_1)\bigl(\langle W\rangle-W_k\bigr), 
\label{eq:SLGC_het}\\ 
W_k &= r_k \mathrm{e}^{\mathrm{i}\phi_k}, \notag \qquad \langle W\rangle = \frac{1}{N}\sum_{j=1}^{N}W_j, \qquad \omega_k \sim \mathcal{N}(\mu,\sigma^2).
\end{align}

Here, $W_k$ denotes the complex amplitude of oscillator $k$, with amplitude $r_k$ and phase $\phi_k$. The parameter $K>0$ sets the strength of the global diffusive coupling, $C_1$ determines the reactive coupling, and $C_2$ controls the nonlinear frequency pulling. Repulsive coupling can be introduced by a suitable choice of parameters. The intrinsic frequencies $\omega_k$ are drawn from a Gaussian distribution with mean $\mu$ and variance $\sigma^2$. 

We choose the parameters to qualitatively reproduce the central experimental phenomenology. Motivated by parameter regimes in which frequency-cluster states have been reported for globally coupled Stuart--Landau oscillators~\cite{Premalatha.2015}, we set $ C_2 =- 7$~\footnote{With our sign convention, this corresponds to $c=7$ in Ref.~\cite{Premalatha.2015}}. Guided by the bimodal-amplitude state observed experimentally and discussed in Ref.~\cite{Thome.2026}, we fix $C_1=-0.5$ and vary $K$. The mean intrinsic frequency can be eliminated by transforming to a rotating frame; hence, without loss of generality, we set $\mu=0$. We choose a standard deviation of $\sigma=0.3$.

\begin{figure*}[h!]
\centering
\includegraphics[width=1.0\textwidth]{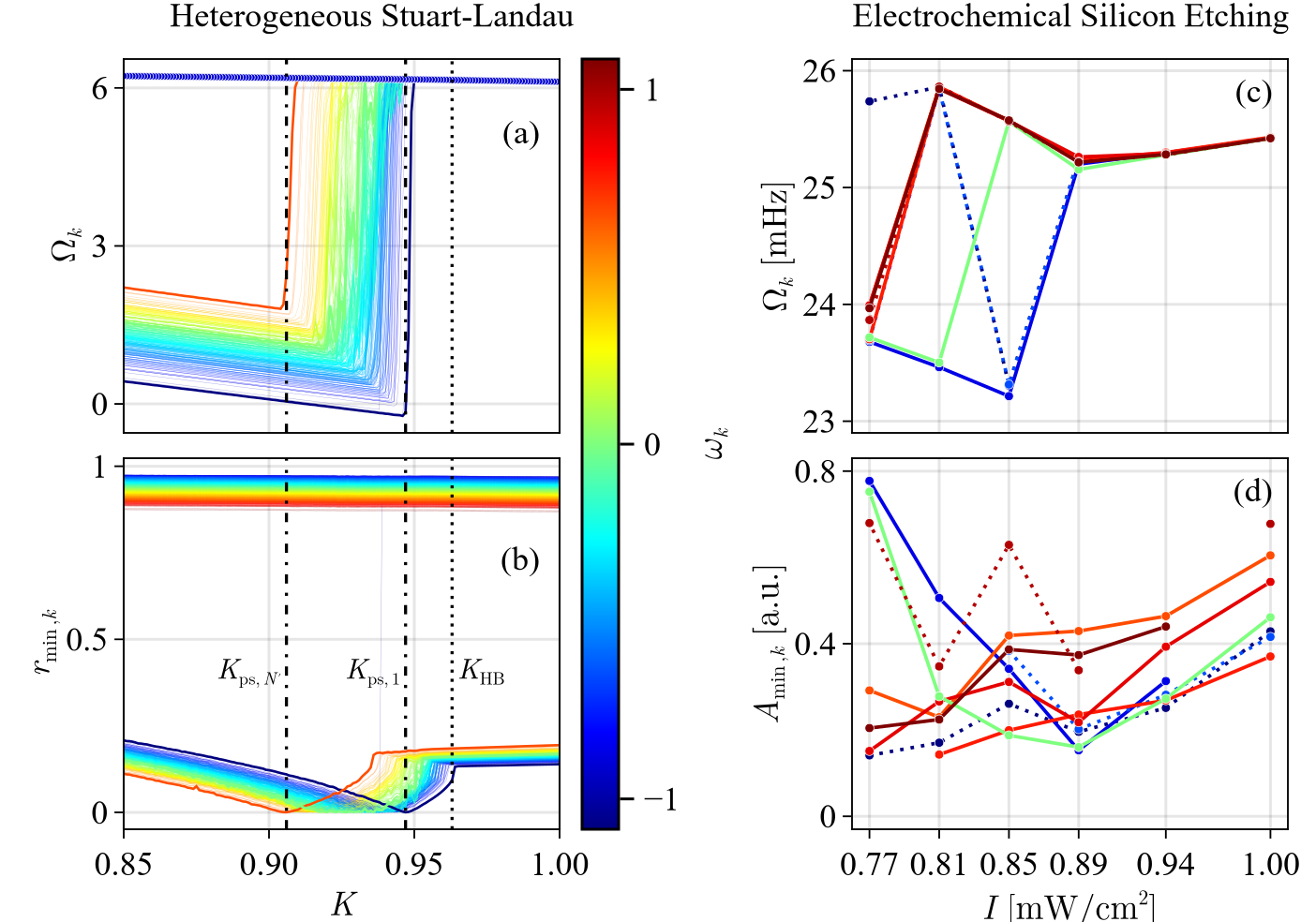}
\caption{\textbf{Transition from coherence to partial incoherence:} In the first two panels, (a) and (b), the numerical results are plotted for a system initialized at a bimodal-amplitude solution for $N=5000$, $C_1=-0.5$, $C_2=-7$, and $ K=1.0$, and adiabatically decreased by steps of $\Delta K=0.001$ until $K=0.85$. The heterogeneity originates from the intrinsic frequencies $\omega_k \sim \mathcal{N}(\mu=0,\sigma=0.3)$ and oscillators are colored according to their $\omega_k$. The top panel in (a) displays the mean frequency $\Omega_k$ of each oscillator $k$, whereas panel (b) displays the minimum of the radius for each oscillator $r_{\text{min, k}}$. The vertical dotted line indicates $K_{\mathrm{HB}}$, where the lowest amplitude oscillator passed the localized Hopf bifurcation. The value $K_{\text{ps},1}$, marked by the dash-dotted line, indicates the first, and $K_{\text{ps},N'}$ the last phase singularity where $r_{\text{min, k}}\approx0$. In panels (c) and (d), we display the results from the electrochemical experiment. Each color corresponds to the analysis result at 9 different local points, and the colors are chosen so that blue oscillators first generate a new frequency, followed by the red oscillators. Dotted lines do not reproduce our results. 
} 
\label{fig:Fig_2}
\end{figure*}

Panels (a) and (b) in Fig.~\ref{fig:Fig_2} show an adiabatic continuation of the bimodal-amplitude state for $N=5000$. Starting from $K=1$, the coupling strength is decreased in steps of $\Delta K=0.001$. Panel (a) shows the mean frequency $\Omega_k$ of each oscillator $k$ (as defined in eq.~\eqref{eq:mean_freq}), while panel (b) shows its minimum amplitude $r_{\min,k}$. Colors indicate the corresponding intrinsic frequencies $\omega_k$. At $K=1$, we start from an established bimodal-amplitude solution of the globally coupled system, whose emergence and properties have been investigated in Ref.~\cite{Thome.2026}. Upon decreasing $K$, the first low-amplitude oscillator develops an amplitude modulation at $K_{\mathrm{HB}}$, consistent with a Hopf bifurcation. Further oscillators undergo analogous transitions as $K$ is decreased. These modulations introduce secondary frequencies, but initially leave the common mean frequency unchanged. To verify that these Hopf bifurcations are oscillator-specific rather than collective instabilities of the entire locked population, we computed the linear stability of the locked state in the frame rotating with $\Omega$ (See section \textbf{S1} in the Supplementary Information for details on stability analysis and mode localization). At the critical coupling strength, a complex-conjugate pair of eigenvalues crosses the imaginary axis. The associated critical eigenmode is strongly localized (defined using the participation ratio PR introduced in Ref.~\cite{Bell.1970}) on the oscillator undergoing the transition. The continued frequency-locked fixed-point branch thus undergoes successive Hopf bifurcations with strongly localized critical eigenmodes. These instabilities are consistent with the sequential onset of amplitude modulation observed in the numerical simulations. Additionally, we confirm that the oscillator with the lowest amplitude is the first one to pass a Hopf bifurcation. 

At a lower, oscillator-dependent coupling strength $K_{\mathrm{ps},1}$, the first modulated trajectory reaches the origin, such that $r_{\min,k}=0$. At this instant, the phase $\phi_k$ is undefined. We refer to such an event as a phase singularity~\cite{Winfree.2001, Aranson.2002}. For each oscillator in the low-amplitude group, the occurrence of a phase singularity coincides with a transition from the common locked frequency to a lower mean frequency. The highlighted dark blue and orange lines illustrate the first and last transitions, respectively. The last transition occurs at $K_{\mathrm{ps}, N'}$, where $N'$ corresponds to the number of oscillators in the low-amplitude cluster. After these successive singularities, the high-amplitude oscillators remain frequency-locked, whereas the low-amplitude group exhibits a broad distribution of mean frequencies.

Panels (c) and (d) of Fig.~\ref{fig:Fig_2} present the behavior of nine representative local signals for the experimental silicon system at different illumination conditions. The minimum amplitude $A_{\mathrm{min,\,k}}$ in panel (d) was extracted by fitting the temporal evolution of the amplitudes of the analytic signals with a sinusoidal function; two examples are provided in the Supplementary Information, section \textbf{S2}. 
For some oscillators and illumination conditions, the fit quality was insufficient to determine $A_{\mathrm{min,\,k}}$, and these data points are not plotted (e.g., blue solid line at $I=1.00~\mathrm{mW/cm^2}$, or only seven successful fits at $I=0.77~\mathrm{mW/cm^2}$). Different colors indicate data at different local positions. The color coding was assigned such that frequency shifts to new modes are broadly grouped into green, blue, and red categories. Panel (c) displays the mean frequency transition as a function of illumination intensity.
When illumination intensity decreases, the blue oscillators transition to a lower frequency between $0.89$ and $0.85~\mathrm{mW/cm^2}$, followed sequentially by the green and then the red oscillator groups.

Panel (d) shows the corresponding transition in $A_{\mathrm{min,\,k}}$ as the illumination intensity decreases. First, we discuss the trend of the data represented by solid lines. As the illumination intensity decreases, the oscillators' minimum amplitudes approach their minima before the frequency transition. After transitioning to lower frequencies, the blue and green oscillators return to higher amplitudes. For the green oscillator, the minimum amplitude increases slightly before the frequency transition occurs.

These findings are consistent with the numerical scenario: oscillators with low amplitudes shift to lower frequencies first, and once a transition to a lower-frequency cluster occurs, the amplitude minimum increases again. The dotted data, however, deviate from this trend, and we return to this point in the discussion. 

\begin{figure*}[h!]
\centering
\includegraphics[width=1.0\textwidth]{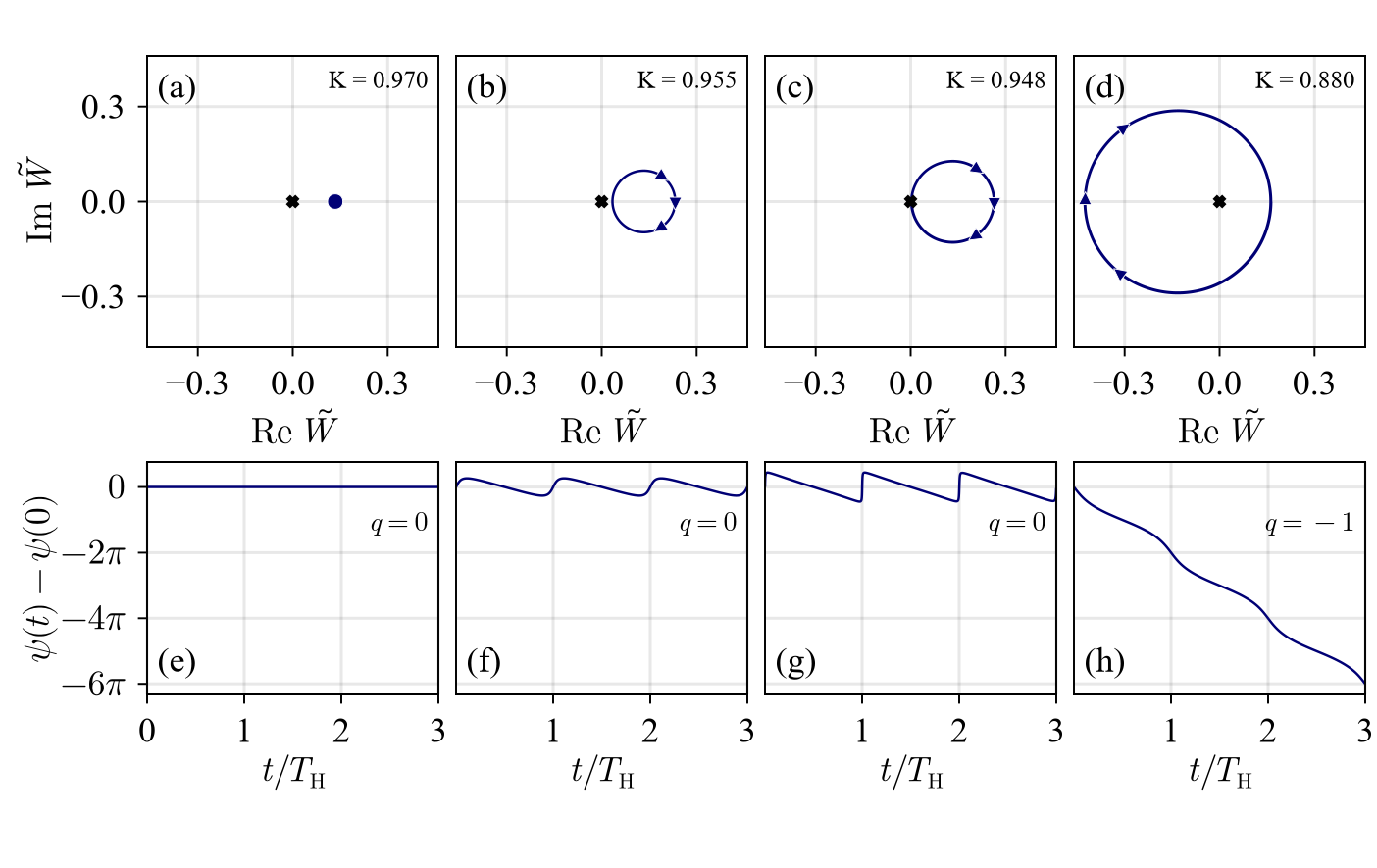}
\caption{\textbf{Representative complex plane representations:} Four representative complex plane representations of the low-amplitude, dark-blue highlighted oscillator from Fig.~\ref{fig:Fig_2}. We plot the trajectories of the oscillator's complex amplitude $\widetilde{W}$ in the comoving frame $\Omega$ for three Hopf periods $T_{\mathrm{H}}$. Additionally, we plot the time series of the comoving phase $\psi$ and indicate the temporal winding number $q$. The $K$-values are $K = 0.970$ (a) and (e), $K = 0.955$ (b) and (f), $K = 0.948$ (c) and (g) and $K = 0.880$ in (d) and (h).}
\label{fig:Fig_3}
\end{figure*}

To gain intuition for the transition to the new frequency, we shift the system into the rotating frame with the locked frequency $\Omega^{\mathrm{lock}}$ (as defined in eq.~\eqref{eq:mean_freq_lock}):
\begin{equation}
\label{eq:comoving_frame}
    \widetilde W_k(t)
    =
    W_k(t)\mathrm{e}^{-\mathrm{i}\Omega^{\mathrm{lock}}t}
    =
    r_k(t)\mathrm{e}^{\mathrm{i}\psi_k(t)},
    \qquad
    \psi_k(t)=\phi_k(t)-\Omega^{\mathrm{lock}} t.
\end{equation}
Figure~\ref{fig:Fig_3} shows the rotating-frame trajectory (top row) and relative phase (bottom row) of the dark blue highlighted low-amplitude oscillator in Fig.~\ref{fig:Fig_2} at values of the coupling strength. Before the Hopf bifurcation, the system can be represented as a fixed point, using the transformation from eq.~\eqref{eq:comoving_frame}, displayed in (a) and (e). The oscillator then passes the localized Hopf bifurcation, and a modulation is observed in panel (b). The modulation orbit does not enclose the origin; consequently, its temporal winding number (cf. eq.~\eqref{eq:winding_number}) is $q=0$, and the relative phase remains bounded (panel (f)). Although the Hopf bifurcation introduces an additional timescale, the mean frequency $\Omega_k$ remains at the locked frequency $\Omega^{\mathrm{lock}}$. As $K$ decreases, the modulation orbit expands until it reaches the origin in (c). After passing the singularity, the trajectory encloses the origin clockwise in (d), corresponding to $q_k=-1$. The relative phase then decreases by $2\pi$ during every Hopf period in (h). The mean angular frequency of the oscillator $k$ follows directly from the phase accumulated over one modulation period:

\begin{align}
\Omega_k&=\Omega^{\mathrm{lock}}+\frac{\psi_k(t+T_{\mathrm{H},k})-\psi_k(t)}{T_{\mathrm{H},k}}=\Omega^{\mathrm{lock}}+q_k\omega_{\mathrm{H},k}.
\label{eq:mean_freq_Hopf}
\end{align}

Before the phase singularity emerges, $q_k=0$ and hence $\Omega_k=\Omega^{\mathrm{lock}}$. After the clockwise winding transition, $q_k=-1$, yielding $\Omega_k=\Omega^{\mathrm{lock}}-\omega_{\mathrm{H},k}$. 

\begin{figure*}[h!]
\centering
\includegraphics[width=1.0\textwidth]{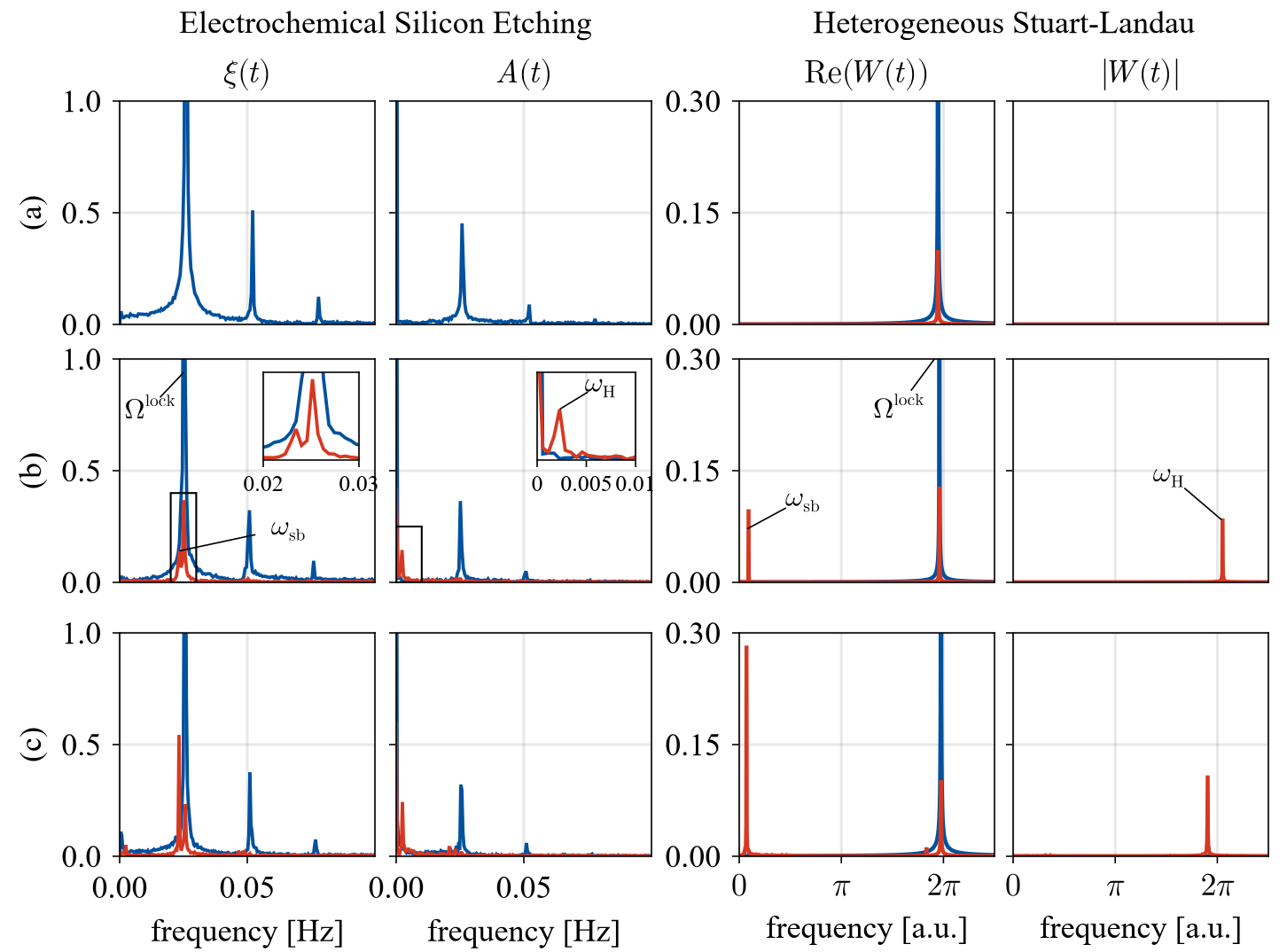}
\caption{\textbf{Fast Fourier Transformation:} One-sided fast Fourier transform for three representative values of the illumination intensity and the coupling strength, respectively. The two columns on the left represent the experimental results, whereas the two columns on the right give the numerical results. Blue data corresponds to an oscillator with high amplitude, and red to one with low amplitude. In panel (a), the illumination intensity is set to $I=3.13~\mathrm{mW/cm^2}$, and the coupling strength is set to $K = 1.0$. In (b), $I=0.94~\mathrm{mW/cm^2}, K = 0.955$, and we labelled the mean locked frequency with $\Omega$, the Hopf frequency with $\omega_{\mathrm{H}}$ and the sideband peak by $\omega_{\text{sb}}$. In panel, (c) $I=0.85~\mathrm{mW/cm^2}$ and $ K = 0.88$. }
\label{fig:Fig_4}
\end{figure*}

The rotating-frame representation predicts that the phase singularity reduces the mean angular frequency by exactly the Hopf frequency. To confirm this mechanism in the experiment, we return to the non-rotating laboratory reference frame and analyze the Fourier spectra of the oscillatory signal $\xi(t)$ ($\mathrm{Re}(W(t))$). Additionally, we include the analysis of the amplitude variable to isolate the Hopf frequency $\omega_{\mathrm{H}}$ as elaborated in Supplementary Information~\textbf{S3} using the Jacobi-Anger expansion~\cite{Abramowitz.1965}. The results are presented in Fig.~\ref{fig:Fig_4}, where three distinct dynamical regimes, controlled experimentally by $I$ and numerically by $K$, are shown. A representative oscillator from the high-amplitude cluster is shown in blue, while a representative oscillator from the low-amplitude cluster is shown in red. For the experimental FFT, the oscillators highlighted by circles in Fig.~\ref{fig:Fig_1} for (a), (b), and (d) are selected. For the numerical FFT, blue and red represent low- and high-amplitude oscillators chosen from Fig.~\ref{fig:Fig_2}.

In the first row, strong couplings $I=3.13~\mathrm{mW/cm^2}$ and $K=1.0$ are chosen, where both systems settle on a completely frequency-locked solution. In the experimental results, a dominant frequency peak $\Omega^{\mathrm{lock}}$ is observed together with higher harmonics (at frequencies $m\Omega^{\mathrm{lock}}, \, m\in \mathbb{N}_{>1}$) in the EMSI signal. These higher harmonics are also observable in the amplitude of the Hilbert transform, indicating the slightly relaxational experimental oscillations~\cite{Patzauer.2021}. For the heterogeneous Stuart--Landau system, there is only one frequency, and no peak appears in the amplitude spectrum because the signal is purely harmonic. For intermediate interaction, in (b), $I=0.94~\mathrm{mW/cm^2}$ and $K=0.955$, the low-amplitude oscillator has already passed the Hopf bifurcation, whereas the high-amplitude oscillator has not. Starting with the FFT of $|W(t)|$, one new frequency $\omega_{\mathrm{H}}$ appears for the low-amplitude cluster, which corresponds to the Hopf frequency. In the FFT spectrum of $\mathrm{Re}(W(t))$, the blue spectrum still shows only one peak, $\Omega^{\mathrm{lock}}$, whereas the low-amplitude cluster displays two peaks: the initial locked frequency $\Omega^{\mathrm{lock}}$ and the side-band frequency $\omega_{\mathrm{sb}}$. Studying the exact numerical values clarifies that $\omega_{\mathrm{sb}}=|\Omega^{\mathrm{lock}}-\omega_{\mathrm{H}}|$ (c.f Table 1 in the Supplementary Information \textbf{S4}). It is also possible to demonstrate the relationship between the peak of the Hopf bifurcation in the FFT of $|W(t)|$ and resulting peaks in the FFT of $\mathrm{Re}(W(t))$ analytically (see section \textbf{S3} in the Supplementary Information). The experimental spectra display the same characteristic sequence of spectral changes. In the FFT of the amplitude $A(t)$, the blue peaks stay similar to the spectrum of (a), but there is a peak for the low-amplitude cluster, which corresponds to the Hopf frequency $\omega_{\mathrm{H}}$. Comparing this to the spectrum in the EMSI signal, we keep the dominant peak at $\Omega^{\mathrm{lock}}$ and identify a novel peak at $\omega_{\mathrm{sb}}$. This novel peak lies very close to $|\Omega^{\mathrm{lock}}-\omega_{\mathrm{H}}|$. In the EMSI signal, we thus observe two frequencies: one from the locked frequency $\Omega^{\mathrm{lock}}$ and a sideband $|\Omega^{\mathrm{lock}}-\omega_{\mathrm{H}}|$. In the last row (c), the height of the novel peak has increased and surpassed that of the $\Omega^{\mathrm{lock}}$-peak, indicating that its frequency is now the dominant one. For the high-amplitude cluster, the dominant frequency lies at $\Omega^{\mathrm{lock}}$, whereas the dominant frequency is $|\Omega^{\mathrm{lock}}-\omega_{\mathrm{H},k}|$ for the low-amplitude cluster in both the electrochemical and numerical experiments.

\paragraph{Phase singularities as a route to partial incoherence}

We summarize these results by proposing an amplitude-mediated transition from complete to partial frequency coherence. The scenario starts from a frequency-locked state in which the oscillator amplitudes form a bimodal distribution, with all oscillators sharing the common frequency $\Omega^{\mathrm{lock}}$. As the coupling strength is decreased, the oscillators with the lowest amplitudes successively undergo Hopf bifurcations. Each bifurcation introduces an oscillator-dependent modulation frequency $\omega_{\mathrm{H},k}$, which varies with the intrinsic frequency $\omega_k$, and generates a sideband at $\Omega^{\mathrm{lock}}-\omega_{\mathrm{H},k}$. Immediately after the Hopf bifurcation, the temporal winding number remains $q_k=0$, such that the mean angular frequency is unchanged, $\Omega_k=\Omega^{\mathrm{lock}}$, and the original locked-frequency component remains dominant. Upon further decreasing the coupling strength, the modulation amplitude grows until the trajectory reaches the origin. At this singularity, the phase becomes undefined, and the temporal winding number can change. After the singularity, the rotating-frame trajectory encloses the origin clockwise, corresponding to $q_k=-1$, and the mean angular frequency becomes $\Omega_k=\Omega^{\mathrm{lock}}-\omega_{\mathrm{H},k}$. Consistently, the sideband at $|\Omega^{\mathrm{lock}}-\omega_{\mathrm{H},k}|$, which was already introduced by the Hopf bifurcation, becomes the dominant component in the one-sided Fourier spectrum. Because the low-amplitude oscillators undergo these transitions successively and possess different modulation frequencies, they acquire different mean frequencies, whereas the high-amplitude oscillators remain locked at $\Omega^{\mathrm{lock}}$. The phase singularity transforms the initially frequency-coherent state into a partially frequency-locked state.

\begin{figure*}[h!]
\centering
\includegraphics[width=1.0\textwidth]{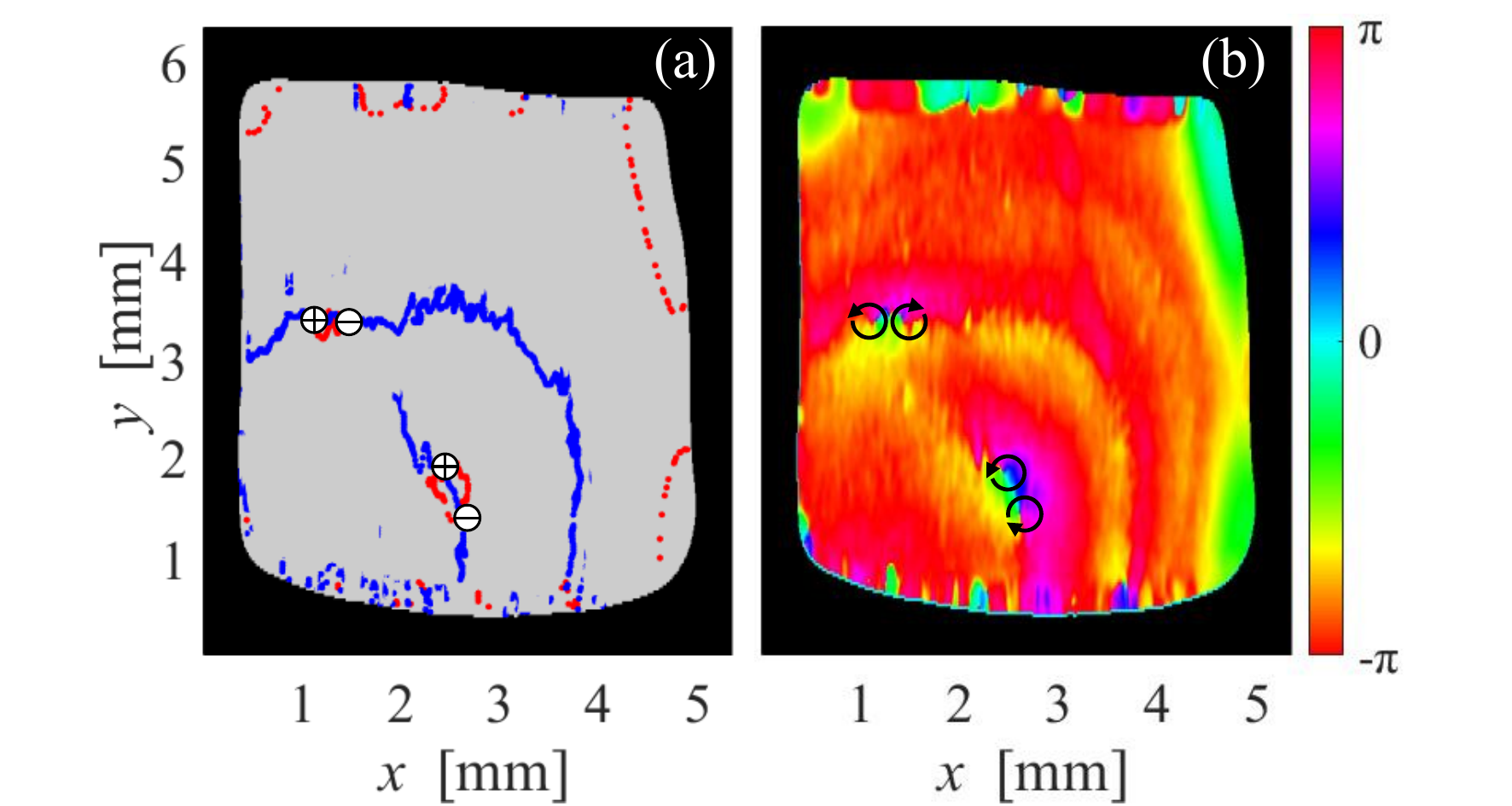}
\caption{\textbf{Topological defect}
Different representation of the experimental results for $I=0.89~\mathrm{mW}/\mathrm{cm }^2$. In panel~(a), we plot the contour of $\mathrm{Re}(\zeta(\vec{x},t'))=0$ in red and of $\mathrm{Im}(\zeta(\vec{x},t'))=0$ in blue, where $t'$ represents an instance of time. The signs $\oplus$ and $\ominus$ represent positive and negative topological charges, respectively. In~(b), a snapshot of the spatial phase field is plotted with arrows pointing to increasing phase values.
}
\label{fig:Fig_5}
\end{figure*}

\paragraph{From temporal phase singularities to spatial topological defects}
In the coupled Stuart--Landau system, the transition to partial incoherence involves a self-organized amplitude distribution, amplitude modulations, a phase singularity, and a change in the spectral profile. While the experiments reproduce the amplitude and spectral signatures, direct evidence of a phase singularity has not yet been presented. Spatial coupling allows defects to migrate, complicating the identification of the phase singularity using the temporal winding-number criterion in eq.~\eqref{eq:winding_number}. Figure~\ref{fig:Fig_1}~(c) shows hints of a momentary phase singularity at isolated spatial locations, where the phase circles around $2\pi$. To study whether these points are actually phase singularities, we study a complementary representation of the experimental data for $I=0.89~\mathrm{mW}/\mathrm{cm }^2$ in Fig.~\ref{fig:Fig_5}. In panel~(a), we plot the contours of $\mathrm{Re}(\zeta(\vec{x},t'))=0$ in red and of $\mathrm{Im}(\zeta(\vec{x},t'))=0$ in blue, where $t'$ represents the time of the phase snapshot in Fig~\ref{fig:Fig_1}. Both lines cross at four distinct spatial points. Thus, at these points, the amplitudes are zero, and the phases are undefined, both characteristics of a phase singularity. In panel~(b), we plot a snapshot of the spatial phase distribution $\phi(\vec{x},t')$. The arrows indicate the direction of increasing phase. Using eq.~\eqref{eq:spatial_winding_number}, we compute the charges of the phase defects. In panel~(a), the signs $\oplus$ and $\ominus$ represent positive and negative topological charges, respectively.
In both systems, we observe phase singularities associated with a change in mean frequency. We used the temporal winding number (eq.~\eqref{eq:winding_number}) for the ensemble of Stuart--Landau oscillators. For the electrochemical experiment, we argue that a non-zero spatial winding number implies a phase singularity. Therefore, in summary, we point out the close correspondence between experiment and theory. 

\begin{figure*}[h!]
\centering
\includegraphics[width=0.5\textwidth]{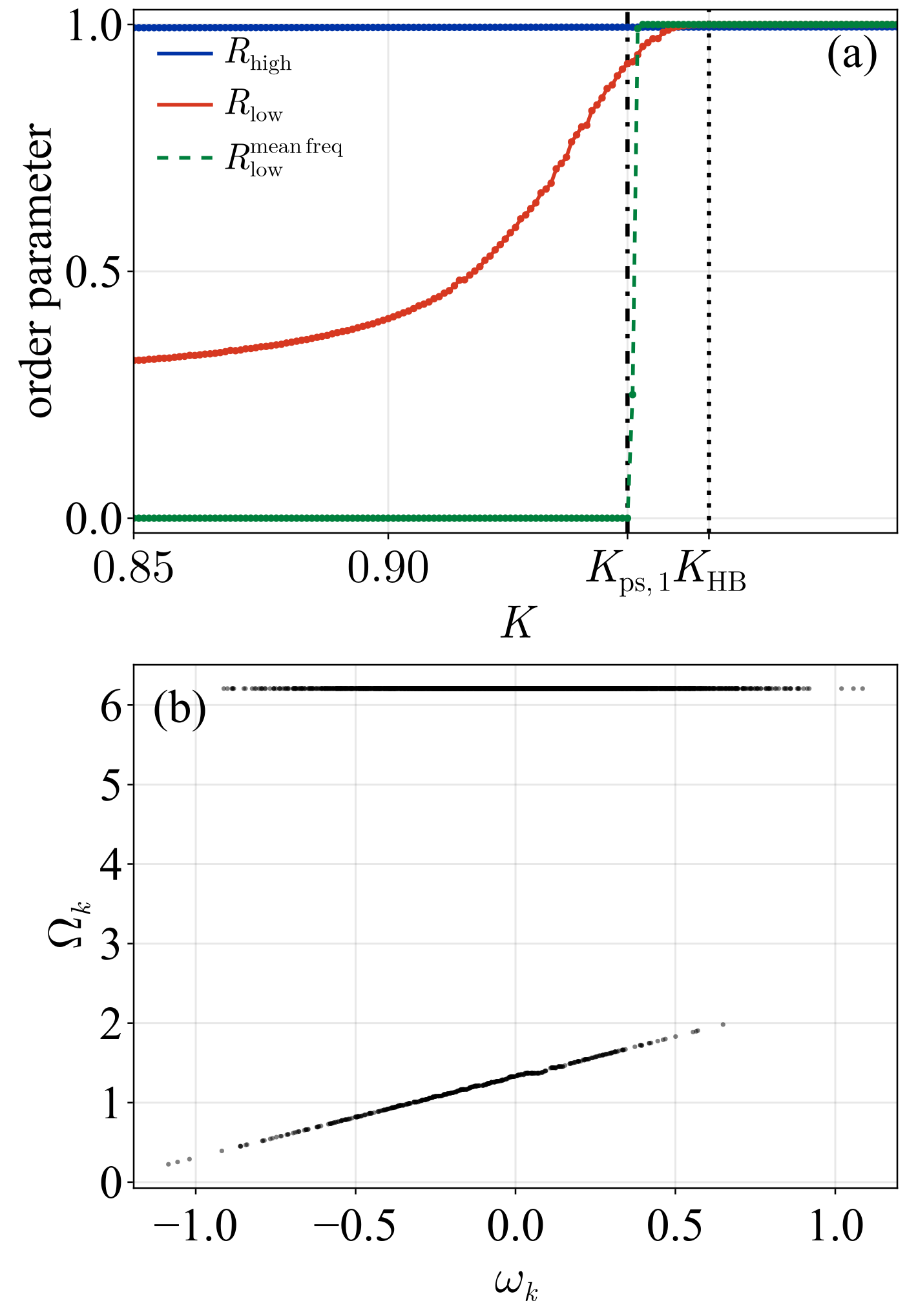}
\caption{\textbf{Macroscopic Transition:} In part (a), we plot different order parameters for the adiabatic continuation plotted in Fig.~\ref{fig:Fig_2}. We plot the Kuramoto order parameter for two subgroups: high-amplitude (blue) and low-amplitude oscillators (red). Additionally, we plot the mean-frequency order parameter for the low-amplitude oscillators in green. In panel (b), we plot the distribution of the mean frequencies $\Omega_k$ against the intrinsic frequencies $\omega_k$ for $K=0.88$.}
\label{fig:Fig_6}
\end{figure*}

\paragraph{Macroscopic description of the transition}
To characterize the amplitude-mediated transition at the population level, phase- and frequency-based order parameters (defined in eqs.~\eqref{eq:KOP} and~\eqref{eq:KOP_mean_Freq}) are evaluated along the adiabatic continuation shown in Fig.~\ref{fig:Fig_2}. The population is divided into high- and low-amplitude subgroups, and the corresponding Kuramoto order parameters $R_{\mathrm{high}}$ and $R_{\mathrm{low}}$ are calculated. Additionally, $R_{\mathrm{low}}^{\text{mean freq}}$ quantifies frequency locking within the low-amplitude subgroup. Figure~\ref{fig:Fig_6}(a) should be interpreted from right to left as $K$ decreases. The high-amplitude subgroup maintains near-perfect phase coherence throughout the transition, with $R_{\mathrm{high}}\simeq1$. The low-amplitude subgroup is also coherent prior to the first localized Hopf bifurcation. At $K=K_{\mathrm{HB}}$, oscillator-dependent amplitude and phase modulations cause $R_{\mathrm{low}}$ to decrease continuously, even though the subgroup remains locked to its mean frequency, characterized by $R_{\mathrm{low}}^{\text{mean freq}}\simeq1$. Consequently, the Hopf bifurcations reduce instantaneous phase coherence without immediately disrupting mean frequency locking. At the first phase singularity, $K=K_{\mathrm{ps,1 }}$, the frequency order parameter decreases abruptly, indicating the onset of oscillator-dependent mean frequencies. The low-amplitude phase-order-parameter remains finite below this transition, suggesting that the drifting oscillators are not uniformly distributed in phase. Their rotating-frame motion exhibits regions of varying angular velocity, which leads to preferential phase accumulation and a nonvanishing subgroup order parameter. The resulting mean-frequency distribution is presented in Fig.~\ref{fig:Fig_6}(b) for $K=0.88$. The high-amplitude oscillators remain locked on the horizontal branch $\Omega_k=\Omega^{\mathrm{lock}}$, while the low-amplitude oscillators acquire distinct mean frequencies that depend on their intrinsic frequencies $\Omega^{\mathrm{lock}}-\omega_{H,k}$. The post-singularity state is therefore partially frequency-locked: the high-amplitude population remains synchronized, whereas the low-amplitude population forms a dispersed frequency branch.

\section{\label{sec:Discussion}Discussion}
We have identified an amplitude-mediated transition from complete frequency coherence to partial incoherence in both an electrochemical oscillator system and a population of globally coupled heterogeneous Stuart--Landau oscillators. This transition originates from a bimodal-amplitude solution and progresses through a sequence of localized Hopf bifurcations and phase singularities. In the Stuart--Landau system, the transition to a different mean frequency is associated with a nonzero temporal winding number, whereas in the electrochemical experiment, it is associated with a nonzero spatial winding number. Oscillators with low amplitudes change their mean frequencies at the phase singularity, whereas high-amplitude oscillators remain frequency-locked at $\Omega^{\mathrm{lock}}$.

This mechanism extends beyond the limitations of a phase-only description. In phase-oscillator models, such as the Kuramoto model, oscillators are restricted to a fixed limit cycle, and the loss of synchronization is indicated by an unbounded phase drift. In contrast, amplitude oscillators possess an additional degree of freedom due to radial dynamics. These oscillators can maintain synchronization by adjusting their amplitudes and can undergo localized Hopf bifurcations while remaining locked to the system's mean frequency. The proposed mechanism combines both: at the phase singularity, the amplitude modulation of a low-amplitude oscillator, initially created by the Hopf bifurcation, becomes so large that the oscillator's trajectory crosses the origin in the complex-amplitude plane. This crossing renders the phase undefined and allows the temporal winding number to change. Therefore, this process transforms a bounded modulation in the rotating frame into an unbounded phase drift, thereby altering the oscillator's mean frequency. 

The phase-singularity-mediated transition reported here is distinct from previously discussed transitions to incoherence in globally coupled amplitude-oscillator systems. Matthews et al.~\cite{Matthews.1991}, for example, treated eq.~\eqref{eq:SLGC_het} for $C_2=C_1=0$. For weak coupling strength and small disorder, the transition is of the Kuramoto-Sakaguchi case. For larger coupling and greater disorder, the authors report a collective Hopf bifurcation of the synchronized state, introducing macroscopic oscillations involving the population as a whole. In contrast, the Hopf bifurcations observed here are localized. On the other hand, related two-stage transitions from phase entrainment through amplitude-modulated states to phase drift have been studied in minimal systems of two coupled amplitude oscillators~\cite{Chakraborty.1988, Aronson.1990}. Here, we extend this scenario to a large, heterogeneous population of ($N=5000$) oscillators, in which the sequential passage through low-amplitude oscillators produces a cascade of oscillator-specific phase singularities and the gradual emergence of partial frequency incoherence.

Electrochemical silicon etching serves as an experimental realization of the amplitude-mediated transition to partial incoherence. At high illumination intensity $I$, the system exhibits a frequency-locked state with pronounced amplitude heterogeneity. As $I$ decreases, this state evolves into a self-organized bimodal-amplitude pattern, where distinct spatial regions exhibit different modulation amplitudes and frequencies. Further reduction in illumination intensity causes the low-amplitude regions to shift to a different dominant frequency, resulting in a spatially extended, partially incoherent state. We use a spatial winding number to identify the existence of a phase singularity, connecting the experimental results to the numerical results.


Yet, the vanishing amplitudes associated with the phase singularities are not directly resolved in Fig.~\ref{fig:Fig_2}(d). There are several reasons for that: First, the experimental minimum amplitudes are estimated from sinusoidal fits to the slow amplitude modulation. Since the electrochemical signal contains higher harmonics, these fits do not necessarily capture the instantaneous minima in amplitude. Second, the finite temporal resolution, discrete illumination steps, and experimental noise limit the accuracy of the estimated minima. Third, unlike in the globally coupled Stuart–Landau system, topological defects in the spatially extended experiment can move. Their motion makes it impossible to obtain a minimum average zero amplitude at fixed measurement locations.

The Stuart--Landau model should be regarded as a minimal mechanistic framework rather than a quantitative representation of the electrochemical experiment. Several intentional simplifications constrain direct quantitative comparison. Most importantly, the model omits spatial coupling via diffusion and therefore cannot capture the spatial gradients and interfaces that develop between different amplitude domains in the experiment (see Fig.~\ref{fig:Fig_1} and Ref.~\cite{Thome.2026}). During the illumination sweep, the spatial pattern of low-frequency clusters changes. As a result, the oscillation at a fixed spatial coordinate can switch between high- and low-frequency modes. The omission of spatial coupling likely explains the local deviations observed in panels (c) and (d) of Fig.~\ref{fig:Fig_2}, where experimental dynamics are shaped by spatial organization absent in the globally coupled oscillator model. Also, experimentally observed traveling waves (Fig.~\ref{fig:Fig_1}(c) phase) are not accounted for in our numerical approach. 

Including spatial coupling would be a natural extension of the model, leading to a complex Ginzburg--Landau equation with additional linear global coupling. Reference~\cite{Thome.2026} showed that adding diffusion to the Stuart--Landau model studied here introduces spatial pattern selection and organizes the different amplitude states into spatial domains, with specific domain walls. On the other hand, the CGLE with additional global coupling exhibits traveling-wave solutions that can interact with amplitude defects~\cite{Battogtokh.1997}. It would thus be fascinating to study how our mechanism relates to the well-known phenomenon of turbulence defects~\cite{Kuramoto.1984, Coullet.1989}. The main challenge is to understand how the spatial organization and the associated interface dynamics interact with the localized amplitude-defect mechanism identified here. 

Another difference between the model and the experiment is that the relative sizes of the amplitude groups differ. In the parameter regime where the amplitude-mediated transition is clearly resolved numerically, the high- and low-amplitude populations do not match the experimental proportions. Consequently, direct quantitative comparison of global order parameters is of limited utility; the corresponding experimental order-parameter evolution is presented in Sec.~\textbf{S5} of the Supplementary Information.

The proposed mechanism suggests a potentially general route to partial incoherence in heterogeneous oscillator media. The main prerequisite for this transition is the amplitude degree of freedom. Based on the experimental observations and the Stuart--Landau model, we identify several additional ingredients that favor such a transition. First, a combination of intrinsic heterogeneity and strong global coupling enables a frequency-locked solution with distinct amplitudes. Second, localized secondary instabilities can produce sufficiently large amplitude modulations to enable phase singularities~\footnote{In the present system, the secondary modulation is generated through a localized Hopf bifurcation of the frequency-locked state. We do not expect this specific bifurcation to be a universal prerequisite for phase-singularity-mediated synchronization loss. More generally, other oscillator-localized bifurcations that generate secondary periodic motion in the rotating frame could provide a similar route.}. Since the Stuart--Landau equation is the normal form of a generic Hopf bifurcation, we expect the mechanism to extend to a broad class of oscillatory systems, provided that coupling and heterogeneity generate the necessary ingredients identified above. Its observation in an electrochemical system beyond the immediate vicinity of a Hopf bifurcation supports this expectation and demonstrates that the mechanism is not restricted to the idealized numerical model.

\section{\label{sec:Methods}Methods}
\textit{Experimental Setup:}

The experimental setup comprises an electrochemical cell, an illumination system, and an ellipso-microscope for surface imaging (EMSI) device.
As working electrodes, single-crystalline n-type Si(111) wafers with a resistivity of $1\text{-}10 \;\rm  \Omega cm$ were cut into $7 \times 7 \,\mathrm{mm}^2$ square samples. 
A $200\,\rm nm$ thick aluminum layer was deposited on the back side, followed by annealing at $250\,^\circ \rm C$ for $15$ min, to establish an ohmic contact.
Prior to the experiments, the samples were plasma oxidized in order to remove residual organic contaminants and passivate the surface.
The silicon electrode was mounted in a custom-built PTFE (polytetrafluoroethylene) holder. 
Electrical connection to the sample was achieved using conductive silver paste, while silicone rubber served as the sealing material. 
In all experiments, the exposed electrode area was maintained at $A = 21.46 \,\rm mm^2$.
A platinum wire and a saturated $\rm Hg|Hg_{2}SO_4$ electrode (MSE) were employed as counter and reference electrodes, respectively.
The electrolyte consisted of an aqueous solution containing $60 \,\rm mM$ $\rm NH_4F$ and $142 \,\rm mM$ $\rm H_2SO_4$ in a total volume of $500 \,\rm ml$, resulting in a pH value of $1$ at room temperature.
Before each experimental run, the solution was purged with argon for $20$ min to remove dissolved oxygen.
Throughout the measurements, continuous mixing was provided by magnetic stirring.
Electrochemical control was performed using a VSP potentiostat (BioLogic).
The electrode potential was fixed at $\mathrm{3.5~V~vs~MSE}$, and an external resistor with an area-normalized resistance of $\mathrm{700~\Omega cm^2}$ was connected in series with the working electrode. 

Photoexcitation required for the electrochemical oxidation process was provided by a He–Ne laser operating at $632\,\rm nm$. The illumination intensity was varied stepwise between $3.13$ and $0.77~\mathrm{mW/cm^2}$ using an electronically controlled neutral-density filter. 

During the electrochemical measurements, an EMSI is used to monitor the spatiotemporal evolution of the silicon oxide layer formed at the interface between the working electrode and the electrolyte.
The recorded signal is the light intensity reflected at the interface.
The signal varies with the optical path length, including changes in the oxide's thickness and the dielectric constant at the interface.
The details of the ellipsometric setup are described in Ref.~\cite{Miethe.2009}.

\textit{Data evaluation of the EMSI:}

To eliminate the linear shift of the intrinsic signal, a linear function was fitted to the temporal signal at each spatial position and subsequently subtracted from the corresponding raw data.
The corrected signals were then subjected to further analysis.

Next, the recorded image sequences were first processed using a Gaussian filter. 
The resulting signal, denoted by $\xi(\vec{x},t)_{\rm raw}$ was normalized according to
\begin{equation}
\xi(\vec{x},t)= \left(\xi(\vec{x},t)_{\rm raw}- \overline{\xi(\vec{x},t)_{\rm raw}} \right) \cdot \frac{\braket{\overline{\xi(\vec{x},t)_{\rm raw}}}}{\overline{\xi(\vec{x},t)_{\rm raw}}},
\end{equation}
where the overlines and cornered brackets indicate the temporal and spatial averages, respectively.
This normalization compensates for small intrinsic variations in the local mean intensity across the monitoring field.

Subsequently, the Hilbert transform~\cite{Boashash.1992} of $\xi(\vec{x},t)$ was calculated, yielding $\xi_{\rm H}(\vec{x},t)$.
The analytic signal was then defined as

\begin{equation}
\label{eq:Hilbert_Tr}
\zeta(\vec{x},t) = \xi(\vec{x},t) +{\rm i} \xi_{\rm H}(\vec{x},t) = A(\vec{x}, t){\rm e}^{{\rm i}\phi(\vec{x}, t)},
\end{equation}
where $A(\vec{x}, t)$ and $\phi(\vec{x}, t)$ represent the instantaneous amplitude and phase of the local oscillatory dynamics, respectively.
The local oscillation frequency, $\omega(\vec{x})$, was determined from the slope obtained by linearly fitting the temporal evolution of $\phi(\vec{x},t)$ over a selected time interval. 
In addition, temporal smoothing was applied to the EMSI data to further suppress high-frequency noise.

\textit{Numerical Integration:}
    
Equation~\eqref{eq:SLGC_het} was numerically solved using the packages SciML/OrdinaryDiffEq~\cite{Rackauckas.2017} in \textit{Julia}~\cite{Bezanson.2017}. For the ODE integration, we used the fifth-order Runge--Kutta method \textit{Tsit5}~\cite{Tsitouras.2011} with adaptive internal time steps and absolute and relative tolerances
$\epsilon_{\mathrm{abs}}=\epsilon_{\mathrm{rel}}=10^{-8}$.
For each value of $K$, we discarded an initial transient of
$T_{\mathrm{tr}}=3000$ and computed observables over a subsequent time window of length $T=600$, sampled at intervals $\Delta t=0.1$. The intrinsic frequencies were drawn using a \textit{MersenneTwister}~\cite{Matsumoto.1998} seed $1234$ and symmetrized about $\mu=0$. The initial condition was generated using seed $5678$. To follow stable solutions in parameter space, we used an adiabatic continuation in the coupling strength $K$. After integrating the system at $K_n$, the final state was used as the initial condition for the next value $K_{n+1}=K_n-\Delta K$, without adding noise or external perturbations. 

\textit{Numerical Dynamical Quantities:} 
    
We define the mean frequency $\Omega_k$ of oscillator $k$ over the measurement window as
\begin{equation}
    \label{eq:mean_freq}
    \Omega_k
    =
    \frac{
    \phi_k(T_{\mathrm{tr}}+T)-\phi_k(T_{\mathrm{tr}})
    }{T},
\end{equation}
where $T_{\mathrm{tr}}$ is the discarded transient time and $T$ is the duration of the subsequent measurement interval. Additionally, we define the locked frequency $\Omega^{\mathrm{lock}}$ as the mean frequency of the high amplitude cluster, 
\begin{equation}
    \label{eq:mean_freq_lock}
    \Omega^{\mathrm{lock}}
    =
    \frac{
    \phi(T_{\mathrm{tr}}+T)-\phi(T_{\mathrm{tr}})
    }{T}.
\end{equation}
To distinguish locked and drifting solutions, we also consider the temporal winding number
\begin{equation}
    \label{eq:winding_number}
    q_k
    =
    \frac{1}{2\pi}
    \int_0^{T_{\mathrm{H},k}} \dot{\psi}_k(t)\,\mathrm{d}t
    \quad k\in \mathbb{Z},
\end{equation}
where $\psi_k$ denotes the phase of oscillator $k$ in the appropriate rotating frame, and $T_{\mathrm{H},k}$ is the period associated with the Hopf frequency. Locked solutions correspond to $q=0$, whereas drifting solutions are characterized by $q \in \mathbb{Z} \backslash \{0\}$. When a spatially extended system is analyzed, it is useful to consider a spatial winding number: 
\begin{equation}
    \label{eq:spatial_winding_number}
    n
    =
    \frac{1}{2\pi}
    \oint_{\gamma} \nabla\phi(\vec{x},t')\cdot\mathrm{d}\vec{x},
\end{equation}
along the contour $\gamma$ which encircles the phase singularity counterclockwise.

We use two types of order parameters. First, we compute the Kuramoto order parameter
\begin{equation}
    \label{eq:KOP}
    R(t)
    =
    \left|
    \frac{1}{N}
    \sum_{k=1}^{N}
    \exp\!\left(\mathrm{i}\phi_k(t)\right)
    \right|.
\end{equation}
In practice, we report its time average over the measurement window. We also compute the same quantity separately for the high- and low-amplitude oscillator groups. These groups are defined by the minimal radius attained during the measurement interval,
\begin{equation}
    r_{\min,k}
    =
    \min_{t\in[T_{\mathrm{tr}},T_{\mathrm{tr}}+T]} r_k(t),
\end{equation}
using a threshold value $r_{\min,\mathrm{c}}$ to separate low-amplitude from high-amplitude oscillators. Second, we quantify frequency locking using a frequency order parameter adapted from~\cite{Newman.2025}. For the mean frequencies, we define
\begin{equation}
    \label{eq:KOP_mean_Freq}
    R^{\mathrm{mean\,freq}}
    =
    \max\left[
    1
    -
    \frac{
    \operatorname{Var}\!\left(
    \Omega_1,\dots,\Omega_N
    \right)
    }{
    \operatorname{Var}\!\left(
    \omega_1,\dots,\omega_N
    \right)
    },
    0
    \right].
\end{equation}
Thus, $R^{\mathrm{mean\,freq}}\approx 1$ indicates that the oscillators have nearly identical mean frequencies, while smaller values indicate frequency dispersion relative to the intrinsic frequency distribution.

\subsection*{Acknowledgements}
The authors thank Reese Caliman for her work on the experimental setup. This project was funded by the Deutsche Forschungsgemeinschaft (DFG, German Research Foundation, project KR1189/18-2).

\subsection*{Declaration of Interests}
The authors declare no competing interests.

\subsection*{Data Availability}
The experimental electrochemical results are stored in a Zenodo repository under accession code ZenodoDOI. New simulation data can be generated from archived GitHub repositories using GitDOI.
\subsection*{Code Availability}
The code used to generate simulation data presented in this study has been deposited in the archived GitHub repositories on Zenodo under accession codes GitDOI and ZenodoDOI.

\subsection*{Author contributions}
N.T. conceived the study. Y.M. designed and performed the experiments and analyzed the experimental data. N.T. developed the numerical framework and performed the corresponding simulations. Y.S. contributed to the interpretation of the results. N.T. and Y.M. wrote the first draft of the main manuscript. N.T. and Y.S. wrote the first draft of the supplementary information. N.T. and Y.S. reviewed and edited the manuscript. K.K. supervised the project, contributed to the interpretation of the results, revised the manuscript, and acquired funding.

\bibliographystyle{unsrtnat}

\clearpage

\setcounter{section}{0}
\setcounter{subsection}{0}
\setcounter{subsubsection}{0}
\setcounter{figure}{0}
\setcounter{table}{0}
\setcounter{equation}{0}
\renewcommand{\thesection}{S\arabic{section}}
\renewcommand{\thesubsection}{S\arabic{section}.\arabic{subsection}}
\renewcommand{\thesubsubsection}{S\arabic{section}.\arabic{subsection}.\arabic{subsubsection}}
\renewcommand{\thefigure}{S\arabic{figure}}
\renewcommand{\thetable}{S\arabic{table}}
\renewcommand{\theequation}{S\arabic{equation}}

\renewcommand{\theHsection}{supp.\arabic{section}}
\renewcommand{\theHsubsection}{supp.\arabic{section}.\arabic{subsection}}
\renewcommand{\theHsubsubsection}{supp.\arabic{section}.\arabic{subsection}.\arabic{subsubsection}}
\renewcommand{\theHfigure}{supp.\arabic{figure}}
\renewcommand{\theHtable}{supp.\arabic{table}}
\renewcommand{\theHequation}{supp.\arabic{equation}}

\begin{center}
    {\LARGE\bfseries Supplementary Information\par}
    \vspace{0.6em}
    {\Large Phase Singularities as a Route to Partial Incoherence in Coupled Oscillators -- From Experiment to Theory\par}
    \vspace{1.2em}
    {\large Nicolas Thom\'e, Yukiteru Murakami, Yannick Sch\"ohs, and Katharina Krischer\par}
    \vspace{0.5em}
    {\normalsize School of Natural Sciences, Physics Department, Nonequilibrium Chemical Physics,\\
    Technische Universit\"at M\"unchen, James-Franck-Str. 1, D-85748 Garching, Germany\par}
\end{center}
\vspace{1.5em}
\section{Linear stability and localization of the Hopf bifurcations}
\label{sec:Lin_stab_Appendix}

To determine whether the Hopf bifurcations observed upon decreasing $K$ are collective instabilities or are localized on individual oscillators, we compute the linear stability of the frequency-locked solution. In the laboratory frame, a completely frequency-locked state rotates with the common angular frequency. We therefore introduce the rotating-frame variables
\begin{equation}
\widetilde W_k(t)
=
W_k(t)\mathrm{e}^{-\mathrm{i}\Omega t}
=
x_k(t)+\mathrm{i}y_k(t).
\label{eq:supp_rotating_frame}
\end{equation}
In this frame, the locked solution corresponds to a fixed point
\begin{equation}
\widetilde W_k(t)=\widetilde W_k^{*}
=
x_k^{*}+\mathrm{i}y_k^{*}.
\label{eq:supp_fixed_point}
\end{equation}
The rotating-frame dynamics are
\begin{equation}
\partial_t\widetilde W_k
=
\left[1+\mathrm{i}(\omega_k-\Omega)\right]\widetilde W_k
-
(1+\mathrm{i}C_2)|\widetilde W_k|^2\widetilde W_k
+
K(1+\mathrm{i}C_1)
\left(
\langle\widetilde W\rangle-\widetilde W_k
\right).
\end{equation}
We introduce
\begin{equation}
\boldsymbol{X}
=
(x_1,y_1,\ldots,x_N,y_N)^{\mathsf T}
\end{equation}
and denote by $\boldsymbol{F}(\boldsymbol{X};\Omega,K)$ the
$2N$-dimensional real vector field obtained by separating the
rotating-frame equations into their real and imaginary parts such that
\begin{equation}
\dot{\boldsymbol{X}}
=
\boldsymbol{F}(\boldsymbol{X};\Omega,K).
\end{equation}
For each value of $K$, the locked state $\boldsymbol{X}^*$ and its
common frequency $\Omega^*$ are determined jointly from
\begin{equation}
\boldsymbol{F}(\boldsymbol{X}^*;\Omega^*,K)
=
\boldsymbol{0},
\qquad
\frac{1}{N}\sum_{k=1}^{N}y_k^*=0.
\end{equation}

The second equation fixes the arbitrary global phase. Thus, the unknown vector of the nonlinear problem is
\begin{equation}
\boldsymbol{q}^*=(x_1^*,y_1^*,\ldots,x_N^*,y_N^*,\Omega^*)^{\mathsf T}.
\end{equation}
The stability of the locked state is governed by the $2N\times2N$ Jacobian

\begin{equation}
\boldsymbol{J}=\left.
\frac{\partial\boldsymbol{F}(\boldsymbol{X};\Omega^*)}
{\partial\boldsymbol{X}}
\right|_{\boldsymbol{X}=\boldsymbol{X}^{*}}.
\label{eq:supp_jacobian}
\end{equation}

For completeness, the Jacobian is written in terms of $2\times2$ blocks as
 \begin{equation}
 \boldsymbol{J}_{k\ell}=\delta_{k\ell}\boldsymbol{A}_k+\frac{K}{N}\boldsymbol{G},
\qquad k,\ell=1,\ldots,N,
 \label{eq:supp_jacobian_blocks}
 \end{equation}

where $\boldsymbol{J}_{k\ell}$ denotes the $(k,\ell)$th block of the Jacobian with

\begin{equation}
\boldsymbol{G}=
\begin{pmatrix}
1 & -C_1\\
C_1 & 1
\end{pmatrix},
\label{eq:supp_global_block}
\end{equation}
and
\begin{equation}
\boldsymbol{A}_k
=
\begin{pmatrix}
a_k-2(x_k^{*})^2+2C_2x_k^{*}y_k^{*}
&
-b_k-2x_k^{*}y_k^{*}+2C_2(y_k^{*})^2
\\
b_k-2C_2(x_k^{*})^2-2x_k^{*}y_k^{*}
&
a_k-2(y_k^{*})^2-2C_2x_k^{*}y_k^{*}
\end{pmatrix},
\label{eq:supp_local_block}
\end{equation}
with
\begin{equation}
a_k=1-K-(x_k^*)^2-(y_k^*)^2,
\qquad
b_k=(\omega_k-\Omega^*)-KC_1
-C_2\left[(x_k^*)^2+(y_k^*)^2\right].
\end{equation}

Owing to the global phase-rotation symmetry, the Jacobian possesses a neutral eigenmode with eigenvalue zero. Its tangent vector is
\begin{equation}
\boldsymbol{v}_{\mathrm{ph}}
=
(-y_1^*,x_1^*,\ldots,-y_N^*,x_N^*)^{\mathsf T}.
\end{equation}
We identify the eigenvector with the largest normalized overlap with
$\boldsymbol{v}_{\mathrm{ph}}$ and exclude this neutral mode when selecting the leading stability eigenvalue. A Hopf bifurcation occurs when a complex-conjugate pair of eigenvalues
\begin{equation}
\lambda_{\mathrm{c},\pm}
=
\alpha_{\mathrm{c}}
\pm \mathrm{i}\omega_{\mathrm{H}}
\label{eq:supp_critical_eigenvalues}
\end{equation}
crosses the imaginary axis. At the critical coupling strength $K_{\mathrm{HB}}$,
\begin{equation}
\operatorname{Re}\lambda_{\mathrm{c},\pm}=0,
\qquad
\operatorname{Im}\lambda_{\mathrm{c},\pm}
=
\pm\omega_{\mathrm{H}}\neq0.
\label{eq:supp_hopf_condition}
\end{equation}
The imaginary part of the corresponding critical eigenvalue therefore gives the Hopf frequency in the frame rotating with $\Omega$.

To quantify how many oscillators participate in the critical eigenmode, we consider the eigenvector $\boldsymbol{v}_{\mathrm{c}}$ associated with $\lambda_{\mathrm{c},+}$. The normalized contribution of oscillator $k$ is defined as
\begin{equation}
p_k
=
\frac{
|v_{\mathrm{c},2k-1}|^2
+
|v_{\mathrm{c},2k}|^2
}{
\displaystyle
\sum_{j=1}^{N}
\left(
|v_{\mathrm{c},2j-1}|^2
+
|v_{\mathrm{c},2j}|^2
\right)
},\quad \sum_{k=1}^{N}p_k=1.
\label{eq:supp_mode_weights}
\end{equation}

The corresponding participation ratio is
\begin{equation}
\mathrm{PR}
=
\frac{1}
{\displaystyle\sum_{k=1}^{N}p_k^2}.
\label{eq:supp_participation_ratio}
\end{equation}
A value $\mathrm{PR}=1$ corresponds to an eigenmode supported by a single oscillator, whereas $\mathrm{PR}=N$ corresponds to a mode distributed uniformly over the entire population.

\begin{figure*}[th]
\centering
\includegraphics[width=0.8\textwidth]
{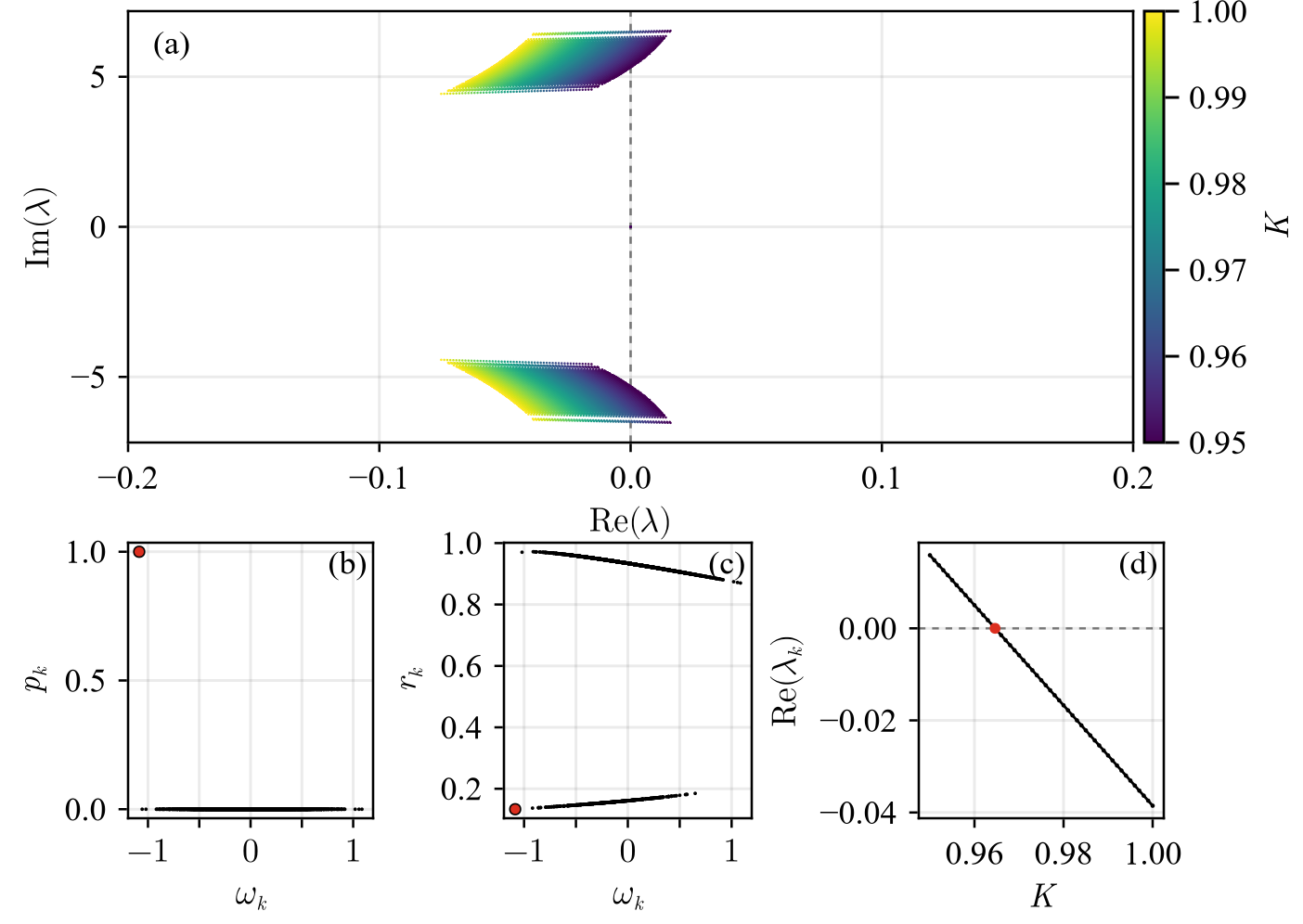}
\caption{
Linear stability and localization of the critical Hopf mode.
(a) Complex-plane representations of all the eigenvalues as a function of $K$. The colorbar indicates the $K$-value, which changes from 1.0 to 0.95 in steps of 0.001. The eigenvalues are computed from fixed-point values in the comoving frame. (b) Oscillator-resolved weights $p_k$ of the corresponding critical eigenvector ($K=K_{\textrm{HB}}$). The mode is strongly localized on a single low-amplitude oscillator, yielding $\mathrm{PR}=1.000194$. (c) Amplitude distribution at the critical bifurcation $K=K_{\mathrm{HB}}$. The colored point is the oscillator $k$ that passes the Hopf bifurcation. (d) Real part of $\lambda_k$ of the critical oscillator $k$ that first passes the Hopf bifurcation. 
}
\label{fig:supp_hopf_localization}
\end{figure*}

Performing the linear stability analysis for the main manuscript continuation from Fig. 2, the first Hopf bifurcation occurs at $K=K_{\text{HB}}$, we obtain a participation ratio of:
\begin{equation}
\mathrm{PR}=1.000194.
\label{eq:supp_pr_result}
\end{equation}
The critical mode is therefore strongly localized on one low-amplitude oscillator, while the contributions of all other oscillators are several orders of magnitude smaller. We plot the evolution of all the eigenvalues in panel (a) of Fig~\ref{fig:supp_hopf_localization}. In panel (b), we plot the oscillator-resolved weights $p_k$ of the corresponding critical eigenvector. In panel (c), we demonstrate that the oscillator with the lowest amplitude corresponds to the critical oscillator. In panel (d), we show the evolution of the leading non-zero eigenvalue. 

Concluding, this demonstrates that the observed instability is not a collective Hopf bifurcation of the entire locked population. Instead, an individual low-amplitude oscillator loses stability while the remaining population stays close to the locked solution. Repeating the analysis at subsequent critical coupling strengths shows that the low-amplitude oscillators destabilize according to their minimal amplitude as $K$ decreases. 

\section{Experimental Data Analysis}
\begin{figure*}[h]
\centering
\includegraphics[width=0.8\textwidth]
{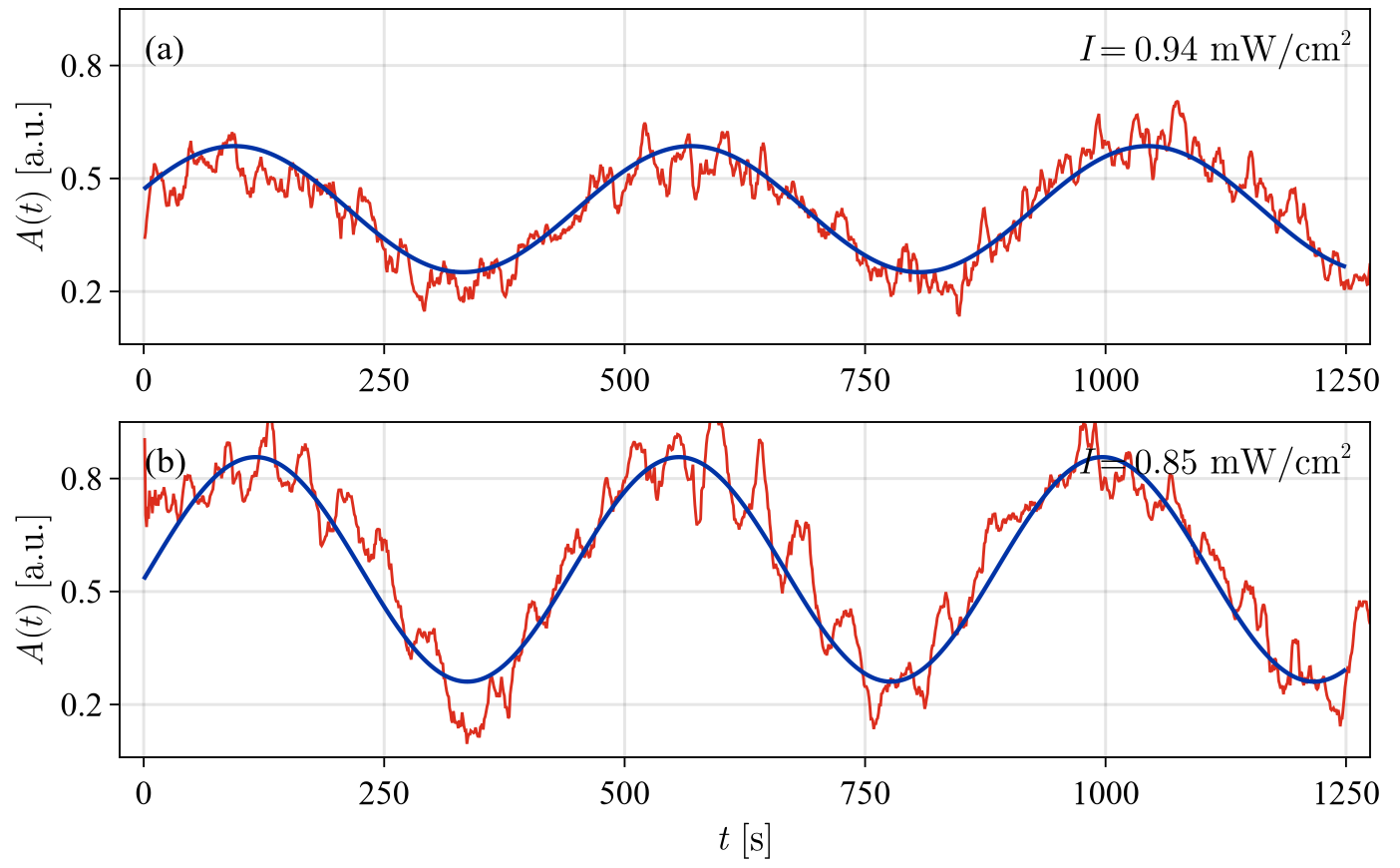}
\caption{
Time series of the amplitude of the local analytic signal for two different illumination values: (a) $I=0.94\,\mathrm{mW/cm^2}$ and (b) $I=0.85\,\mathrm{mW/cm^2}$.
}
\label{fig:TS}
\end{figure*}

We use the minimum amplitude modulation $r_{\mathrm{min}}$, respectively $A_{\mathrm{min}}$, to characterize the phase singularities. 
For the Stuart-Landau model, the numerical computation of $r_{\mathrm{min}}$ is straightforward, since the modulation of $r$ consists of a single frequency. 
In the experimental data, the modulation of the amplitude consists of multiple frequencies. 
To determine the Hopf-induced amplitude modulation, we fit the time evolution of $A(t)$ with a harmonic fit. Here, we want to ignore contributions from the ground oscillation (fast oscillations in Fig~\ref{fig:TS}), which originate from higher harmonics of the locked frequency. 
\begin{equation*}
    A(t)=A_0+y\sin\left(\pi\frac{t-t_c}{w}\right), \qquad A_{\mathrm{min}}=A_0-y
\end{equation*}
Figure~\ref{fig:TS} displays the temporal evolution of the amplitudes of local analytic signals under two different illumination intensities $I$.
The blue curves indicate the sinusoidal fits.
In panel (a), the system is above the threshold of the phase singularity, whereas panel (b) displays the time series after the phase singularity. These two time series correspond to the low-amplitude data from the FFT analysis in panels (b) and (c) from Fig. 2 of the main manuscript. 

\newpage
\section{Relation between the Hopf-frequency in the amplitude variable and the frequency components of the observable $\mathrm{Re}(W(t))$}
In the following, we explain the relation between the Hopf frequency $\omega_{\mathrm{H}}$ of the oscillation amplitude and the frequency components in the observable $\mathrm{Re}(W(t))$.
At first, we remark that for large values of $K$, the oscillators rotate with a common frequency $\Omega$, yielding a limit cycle solution. Switching to the reference frame rotating with frequency $\Omega$ using $\widetilde{\phi}_i=\phi_i - \Omega t$, the cycle is reduced to a fixed point. We have shown in section \ref{sec:Lin_stab_Appendix} that each oscillator undergoes a Hopf bifurcation in this reference frame when decreasing $K$. Sufficiently close after its bifurcation, the dynamics of each oscillator $i$ can therefore, without loss of generality, be approximated by 
\begin{equation}
\label{eq:amplitude}
    r_i(t)=r_{0,i}+ \epsilon_r \sin(\omega_{\mathrm{H}} t),
\end{equation}
\begin{equation}
\label{eq:phase}
    \phi_i(t)=\Omega t + \epsilon_p \cos(\omega_{\mathrm{H}} t),
\end{equation}

where $\epsilon_p$ and $\epsilon_r$ are both much smaller than one.
Using this approximation, we now compute $\mathrm{Re}(W_i(t))$ (we omit the index $i$ in the following) to determine its frequency components in the regime close to the Hopf bifurcation:
\begin{align*}
    \mathrm{Re}(W) &= \mathrm{Re}(r(t) \mathrm{e}^{\mathrm{i} \phi(t)}) =  \\
    &= r(t) \cdot \mathrm{Re}(\mathrm{e}^{\mathrm{i}(\Omega t)} \mathrm{e}^{\mathrm{i}(\epsilon_p \cos(\omega_{\mathrm{H}} t))}) \\ 
    &= r(t) \cdot \mathrm{Re}\left[\mathrm{e}^{\mathrm{i}(\Omega t)} \left(J_0(\epsilon_p) + 2\sum_{n=1}^{\infty} \mathrm{i}^n J_n(\epsilon_p) \cos(n \omega_{\mathrm{H}} t) \right) \right] \\
    &= r(t) \left[J_0(\epsilon_p)  \cos(\Omega t) + 2 \cos(\Omega t) \left(\sum_{n=1}^{\infty} (-1)^n J_{2n}(\epsilon_p) \cos(2n \omega_{\mathrm{H}} t))\right) \right. \\
    &\left. + 2 \sin(\Omega t) \left(\sum_{n=1}^{\infty} (-1)^{n} J_{2n-1}(\epsilon_p) \cos((2n-1) \omega_{\mathrm{H}} t))\right) \right] .
\end{align*}
In the third line, we applied the Jacobi-Anger expansion, where $J_n$ represent the Bessel functions of the first kind. Further simplification of the products of trigonometric functions yields:
\begin{align*}
    \mathrm{Re}(W)&= r(t) \Bigg[ J_0(\epsilon_p) \cos(\Omega t) + \left(\sum_{n=1}^{\infty} (-1)^n J_{2n}(\epsilon_p) \Big[\cos((\Omega + 2n \omega_{\mathrm{H}}) t) + \cos((\Omega - 2n \omega_{\mathrm{H}}) t) \Big] \right) \\ 
    & + \left(\sum_{n=1}^{\infty} (-1)^{n} J_{2n-1}(\epsilon_p) \Big[ \sin((\Omega + (2n-1) \omega_{\mathrm{H}}) t) +\sin((\Omega - (2n-1) \omega_{\mathrm{H}}) t) \Big] \right) \Bigg].
\end{align*}

When further plugging in the amplitude $r(t)$ (eq.~\eqref{eq:amplitude}) and simplifying, we obtain
\begin{align*}
    \mathrm{Re}(W)&= r_0  \Bigg[ J_0(\epsilon_p) \cos(\Omega t) + \left(\sum_{n=1}^{\infty} (-1)^n J_{2n}(\epsilon_p) \Big[\cos((\Omega + 2n \omega_{\mathrm{H}}) t) + \cos((\Omega - 2n \omega_{\mathrm{H}}) t) \Big] \right) \\ 
    & + \left(\sum_{n=1}^{\infty} (-1)^{n} J_{2n-1}(\epsilon_p) \Big[ \sin((\Omega + (2n-1)  \omega_{\mathrm{H}}) t) +\sin((\Omega - (2n-1) \omega_{\mathrm{H}}) t) \Big] \right) \Bigg]\\
    &+ \frac{1}{2} \epsilon_r J_0(\epsilon_p) \Big[\sin((\Omega + \omega_{\mathrm{H}}) t) + \sin((\Omega - \omega_{\mathrm{H}}) t) \Big] \\
    &+ \frac{1}{2} \epsilon_r \left(\sum_{n=1}^{\infty} (-1)^n J_{2n}(\epsilon_p) \Big[\sin((\Omega + (2n-1) \omega_{\mathrm{H}}) t) + \sin((\Omega + (2n+1) \omega_{\mathrm{H}}) t) \right. \\ 
    & \left. + \sin((\Omega - (2n+1) \omega_{\mathrm{H}}) t) + \sin((\Omega - (2n-1) \omega_{\mathrm{H}}) t) \Big] \right. \Bigg) \\
    &{+ \frac{1}{2} \epsilon_r \Bigg(\sum_{n=1}^{\infty} (-1)^{n} J_{2n-1}(\epsilon_p) \Big[\cos((\Omega + (2n-2)\omega_{\mathrm{H}}) t) - \cos((\Omega + 2n \omega_{\mathrm{H}}) t)} \\
    &+ \cos((\Omega - (2n-2) \omega_{\mathrm{H}}) t) - \cos((\Omega - 2n \omega_{\mathrm{H}}) t) \Big] \Bigg).
\end{align*}

Since we are assuming proximity to the Hopf bifurcation, it holds that $\epsilon_r , \epsilon_p \ll 1$. Therefore, we can neglect terms that are of quadratic or of higher order in $\epsilon_i$. For small arguments, the Bessel functions can be approximated as $J_n(\epsilon) = \frac{1}{(n+1)!}\left( \frac{\epsilon}{2} \right)^n + \mathcal{O}(\epsilon^{n+2})$, which yields
\begin{align*}
    \mathrm{Re}(W)&= r_0  \cos(\Omega t) - \frac{1}{4}\epsilon_p r_0 \Big[ \sin((\Omega + \omega_{\mathrm{H}}) t) +\sin((\Omega - \omega_{\mathrm{H}}) t) \Big] \\
    &+ \frac{1}{2} \epsilon_r \Big[\sin((\Omega + \omega_{\mathrm{H}}) t) + \sin((\Omega - \omega_{\mathrm{H}}) t) \Big]
    + \mathcal{O}(\epsilon_p^{2}, \epsilon_p \epsilon_r, \epsilon_r^{2}) .
\end{align*}
The first term represents the rotation at frequency $\Omega$, while the other terms are additional small oscillations due to the Hopf bifurcation at frequencies $\Omega \pm \omega_{\mathrm{H}}$. Thus, we have shown that in the approximation up to the first order, a Hopf bifurcation occurring in amplitude and phase variables only gives rise to additional frequency components of $\Omega \pm \omega_{\mathrm{H}}$ in Cartesian coordinates. Terms oscillating at the Hopf-frequency $\omega_{\mathrm{H}}$ itself do not emerge in $\mathrm{Re}(W)$.

\section{FFT values}
We have plotted the Fast Fourier Spectra for both the electrochemical and numerical experiments. Table~\ref{tab:frequency_comparison}, displays the exact numerical values obtained.
\begin{table*}[th]
\centering
\caption{
Frequencies extracted from the experimental and numerical spectra.
The predicted sideband frequency is
$\omega_{\mathrm{sb}}=|\Omega^{\mathrm{lock}}-\omega_{\mathrm{H}}|$.
}
\label{tab:frequency_comparison}

\begin{tabular}{llcccc}
\hline\hline
System &
Control parameter &
$\Omega^{\mathrm{lock}}$ &
$\omega_{\mathrm{H}}$ &
$\omega_{\mathrm{sb}}$ &
$|\Omega^{\mathrm{lock}}-\omega_{\mathrm{H}}|$ \\
\hline

Experiment
& $I=3.13~\mathrm{mW/cm^2}$
& $0.025789$
& -- & -- & -- \\

& $I=0.94~\mathrm{mW/cm^2}$
& $0.025143$
& $0.002286$
& $0.023429$
& $0.022857$ \\

& $I=0.85~\mathrm{mW/cm^2}$
& $0.025714$
& $0.002449$
& $0.023265$
& $0.023265$ \\

\hline

SLH
& $K=1$
& \angfreq{0.97278}
& -- & -- & -- \\

& $K=0.955$
& \angfreq{0.97993}
& \angfreq{1.025641}
& \angfreq{0.045711}
& \angfreq{0.045711} \\

& $K=0.88$
& \angfreq{0.988501}
& \angfreq{0.952789}
& \angfreq{0.035712}
& \angfreq{0.035712} \\

\hline\hline
\end{tabular}
\end{table*}

The resolution of the FFT peaks is bounded by the length of the time series. For the SLH system, the resolution is $(\pm$ \angfreq{0.000714}$)$ for all values of $K$. 
For the experiment, in the case of $I=0.94 \mathrm{mW/cm^2}$, $\omega_{\mathrm{sb}}=|\Omega^{\mathrm{lock}}-\omega_{\mathrm{H}}|$ is satisfied, while in the case of $I=0.85 \mathrm{mW/cm^2}$, a $\mathrm{0.57~mHz}$ discrepancy is observed, which corresponds to the frequency resolution of the FFT $\Delta f = \mathrm{0.57~mHz}$. 
The observed frequency deviation is attributed to the finite length of the recorded time series.

\section{Kuramoto order parameter for the experiment}
In the experiment, there are a few oscillators in the high-amplitude cluster, whereas there are a few oscillators in the low-amplitude cluster in the numerical experiment. Thus, the two global order parameters characterize different macroscopic solutions, and comparing them does not generate meaningful insight. Here, we plot the order parameter across the entire electrode at different illumination intensity levels $I$. The displayed labels (a), (b), and (c) correspond to the data from Fig.1 of the main document. Since in the experiment only a few oscillators are in the high-amplitude cluster, the global order parameter corresponds roughly to the order parameter for the low-amplitude cluster. The global order parameter of the experiment qualitatively matches the low-amplitude order parameter of the numerics.

\begin{figure*}[th]
\centering
\includegraphics[width=0.5\textwidth]
{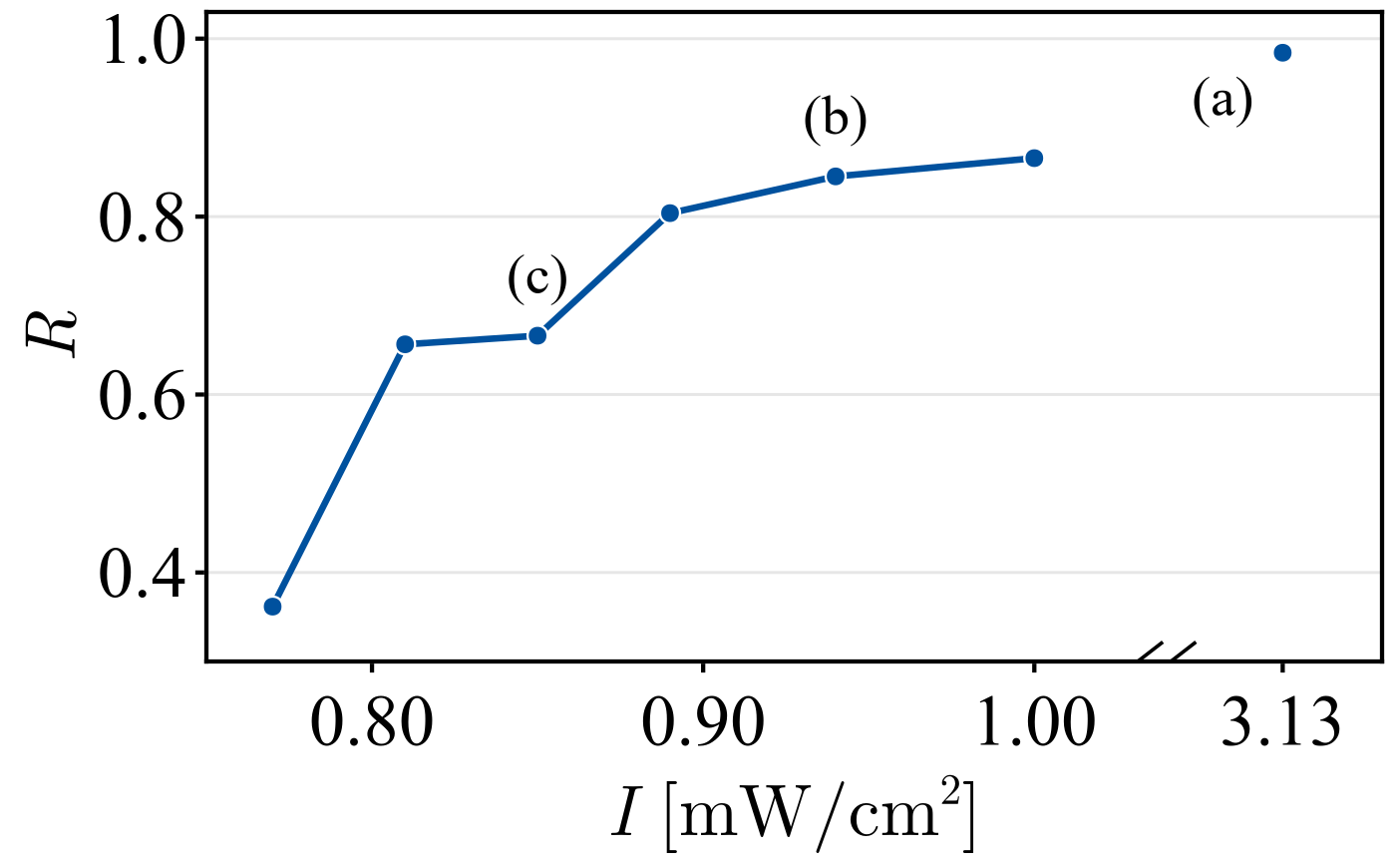}
\caption{
Kuramoto order parameter $R$ as a function of the illumination intensity $I$ for the entire electrode. The labels (a), (b), and (c) correspond to data from Fig. 1 in the main document.
}
\label{fig:Fig_KOP_exp_broken_axis}
\end{figure*}

\end{document}